\documentclass[a4paper,11pt]{article}
\usepackage{graphicx}% Include figure files
\usepackage{dcolumn}% Align table columns on decimal point
\usepackage{bm}% bold math
\usepackage{xcolor}
\usepackage{enumitem}
\usepackage{amsmath}
\usepackage{amsfonts}   
\usepackage{amssymb}
\usepackage{authblk}
\usepackage[utf8]{inputenc}
\usepackage[T1]{fontenc}

\title{Hydrostatic Newton-Cartan Membranes}
\author[a]{Domingo Gallegos \thanks{d.gallegos@ciencias.unam.mx}}
\affil[a]{Universidad Nacional Aut\'onoma de M\'exico, Facultad de Ciencias}
\author[a]{Carlos M\'alaga}
\begin{document}

\maketitle

%\preprint{APS/123-QED}

\begin{abstract}

An application of the Newton-Cartan framework to the study of membranes is presented. Specifically, for membranes of co-dimension one in hydrostatic equilibrium embedded in a flat ambient Newton-Cartan spacetime. For such membranes, the corresponding equilibrium partition function at second order in the hydrodynamic derivative expansion is shown. Equilibrium constraints and the corresponding
set of equilibrium constitutive relations are found. For the generically non-constant elastic subset of thermodynamic coefficients, the Young-Laplace equation is presented for the case of two-dimensional axisymmetric closed membranes embedded in a flat three-dimensional spacetime with constant ambient vorticity. Some numerical solutions to this Young-Laplace equation are examined, and some analytic solutions for particular choices of the thermodynamic coefficients are also discussed.

%\begin{description}
%\end{description}
\end{abstract}

\maketitle

%\tableofcontents

\section{Introduction}\label{sec::intro}

Surfaces and interfaces are of importance in a wide range of physical phenomena such as biological systems \cite{Canham1970,Helfrich,Tu2014,Cowin:2001,Mooendarbary:2013,Mow:1984}, condensed matter systems \cite{Kane2005}, and wave dynamics on spherical surfaces \cite{Streichan2017, Ritsma2014, Delplace2017, Shankar2017, Pearce2019}, such as planet Earth.
Closed surfaces modeling physical shells and membranes in equilibrium can adopt different shapes depending on their free energy minimum.
Typically, elastic contributions to the free energy will lead to spheres, tori, and biconcave discs, similar to the shape observed in red blood cells.

A theoretical framework for studying such systems in equilibrium was established in 2020 by Armas {\it et al.} \cite{Armas:2020}. 
The membrane is treated as a submanifold in a Newton-Cartan spacetime.
A key aspect of the formalism is that a geometric interpretation of temperature and mass chemical potential arises naturally from it, which allows for the study of elastic surfaces at finite temperature and mass chemical potential.
One of their results is a generalization of the Canham-Helfrich free energy for lipid bilayer membranes, which predicts the shapes mentioned above.

In the present manuscript, we make extensive use of this framework to derive general constitutive relations for fluid membranes of co-dimension one in hydrostatic equilibrium embedded in a flat ambient Newton-Cartan spacetime up to second order in the hydrodynamic derivative expansion. 
%%%%%%%
Hydrostatic conditions force the temperature to remain constant while the chemical potential is determined by the rotational kinetic energy of the fluid elements. 
The corresponding free-energy and constitutive relations can be derived, and the variation of such free-energy will lead to the Young-Laplace equation, which will determine the shape of the membrane. The Young-Laplace equation with purely elastic contributions is examined for two-dimensional axisymmetric membranes embedded in a spacetime with constant ambient vorticity. This corresponds to a fluid membrane immersed in a fluid that rotates with constant vorticity. Limiting the constitutive relations to the purely elastic subset results in the Young-Laplace equation depending on the usual elastic properties given by the surface tension, the bending and Gauss moduli, and the spontaneous curvature \cite{Hiroyoshi1996}, but with non-trivial dependence on the chemical potential, which itself will depend on the vorticity through the equilibrium conditions.
New thermodynamic parameters arise through these contributions, the effects of which have not been studied before, although the possibility of including them has been discussed before \cite{Armas:2020, Armas:2013.2}.
Alternatively, part of the present work can be considered as the non-relativistic limit of the framework developed by Armas in \cite{Armas:2013} and \cite{Armas:2013.2}.

This work contributes to the active research on membrane dynamics and demonstrates a simple application of the Newton-Cartan submanifold formalism, showing that hydrostatic membranes can sustain rotation and giving some examples of the shapes they can take in such flows.
We find it worth mentioning the work by Salbreux and Jülicher, who derived, also from a geometrical standpoint, a covariant theory of active surfaces that appear in biological and chemical processes \cite{salbreux}.
Their work provides a framework for studying membranes with microstructures that break symmetries, such as chiral surfaces. 
Symmetry considerations lead them to specific constitutive relations for stress and moment tensors.
For liquid crystal membranes, Vafa and Mahadevan have proposed a framework to study the deformations produced by topological defects \cite{vafa}.
Their work can potentially model epithelial morphogenesis.
The numerical studies of Hoffman, Carenza and Giomi on liquid crystal shells show that intense activity can lead to topological changes \cite{hoffmann}.
In particular, extensile activity can 
transform a flattened sphere into a torus, similarly to the effect of vorticity that was found in some of the numerical solutions shown in this work.

%Surfaces and interfaces are of importance on a wide range of physical phenomena such as biological systems \cite{Canham1970,Helfrich,Tu2014,Cowin:2001,Mooendarbary:2013,Mow:1984}, condensed matter systems \cite{Kane2005}, and wave dynamics on spherical surfaces \cite{Streichan2017, Ritsma2014, Delplace2017, Shankar2017, Pearce2019}, such as planet Earth. A theoretical framework for studying such systems was established in \cite{Armas:2020}, based on the microscopic symmetry structure of the systems under study. A key aspect of the formalism used in \cite{Armas:2020} is that a geometric interpretation of temperature and mass chemical potential arise naturally from it. This allows for the study of elastic surfaces at finite temperature and mass chemical potential.
%%%%%%

%Following the work laid down in \cite{Armas:2020}, it is the goal of this paper to establish the equilibrium states of membranes that allow the exchange of both energy and matter. In particular, we will use a generating equilibrium partition function to write down the most general set of constitutive relations for membranes with finite temperature and chemical potential. 

%%%

The manuscript is organized as follows.
In sections \ref{Sec:TNC} and \ref{Sec:Currents}, a summary of the necessary tools to understand the formalism presented in \cite{Armas:2020} is provided. 
Section \ref{Sec:Eq} shows the use of the formalism to write down the equilibrium partition function up to second order in derivatives together with its corresponding equilibrium constraints and constitutive relations. 
In section \ref{Sec:YL}, we show the corresponding Young-Laplace equation in the presence of constant ambient vorticity for the elastic subset of thermodynamic coefficients assuming a dependence on both temperature and chemical potential. 
Finally, in section \ref{Sec:Con} we conclude with some comments on potential future work. 

\section{Torsional Newton Cartan Background}\label{Sec:TNC}

In this section, we present an overview of Torsional Newton Cartan (TNC) geometry \footnote{Newton-Cartan geometry was first introduced by Cartan in \cite{Cartan:1924,Cartan:1924yea}. A modern understanding of non-relativistic gravity and non-relativistic geometry was pioneered in \cite{Andringa2011}, with more systematic approaches done in works such as \cite{VandenBleeken:2017,Hansen:2019,Hansen:2020}. See \cite{Hartong:2022} for a modern review of Newton-Cartan geometry.}and introduce all the essential details that will be needed when using this geometry as the source of the mass, momentum, and energy currents. The essential details include a summary of TNC sub-manifolds, a formalism that is needed to describe the equilibrium state of non-relativistic membranes. 
The overview presented in this and the next section regarding TNC sub-manifolds can be considered a summary of the formalism of Newton-Cartan submanifolds that was introduced in \cite{Armas:2020}.

\subsection{TNC Geometry}

Let $\mathcal{M}_{d+1}$ be a (d+1)-dimensional manifold equipped with a TNC structure, consisting on the fields\footnote{We use Greek indices to denote space-time indices $\{\mu,\nu,... \}=0,...,d$} $\left(\tau_\mu,h_{\mu \nu},m_\mu \right)$; with $h_{\mu \nu}$ a symmetric spatial metric, $\tau_\mu$ a 1-form known as the clock 1-form, and $m_\mu$ a U(1) gauge connection associated with particle number conservation.  An inverse structure $\left(\upsilon^\mu, h^{\mu \nu} \right)$ can be defined following the orthonormality conditions 
%%%
\begin{align}
	\tau_\mu h^{\mu \nu} &= 0, \qquad \qquad &
	\tau_\mu \upsilon^\mu &=-1, \\
	h_{\mu \nu} \upsilon^\nu &= 0, \qquad \qquad &
	h_{\mu \rho}h^{\rho \nu} - \tau_\mu \upsilon^\nu &= \delta^\nu_\mu.
\end{align}
%%%
The TNC structure on $\mathcal{M}_{d+1}$ in terms of the fields $\left(\tau_\mu,h_{\mu \nu},m_\mu \right)$ transform under diffeomorphisms  (local coordinate transformations parametrized by $\xi^\mu$), a U(1) transformation (mass gauge transformation parametrized by $\sigma$), and Milne boosts  (local Galilean boosts parametrized by $\lambda_\mu$) as 
%%%%%%
\begin{align}
	\delta \tau_\mu &= \mathcal{L}_\xi \tau_\mu \, , \\
	\delta h_{\mu \nu} &= \mathcal{L}_\xi h_{\mu \nu} + \lambda_\mu \tau_\nu + \lambda_\nu \tau_\mu \, , \\
	\delta m_\mu &= \mathcal{L}_\xi m_\mu + \lambda_\mu + \partial_\mu \sigma.
\end{align}
%%%%

A spacetime measure $e$ can be constructed from this objects through the identification
%%%
\begin{align}
	e=\sqrt{ -\det\left(h_{\mu \nu} - \tau_\mu \tau_\nu  \right) }.
\end{align}
%%%
It is sometimes more convenient to work with the explicitly Milne boost invariant structure $\{\tau_\mu, \bar h_{\mu \nu}, \Phi \}$ and inverse structure $\{\hat \upsilon^\mu, h^{\mu \nu} \}$, defined as 
%%%
\begin{align}
	\bar h_{\mu \nu} &\equiv h_{\mu \nu} - \tau_\mu m_\nu - \tau_\nu m_\mu \, , \\
	\hat \upsilon^\mu &\equiv \upsilon^\mu - h^{\mu \nu} m_\nu \, , \\
	\Phi &\equiv - \hat \upsilon^\mu m_\mu + \frac{1}{2} h^{\mu \nu} m_\mu m_\nu \, , 
\end{align}
%%%
and satisfying the orthonormality conditions
%%%
\begin{align}
	\tau_\mu h^{\mu \nu} &= 0 \qquad \qquad & \tau_\mu \hat \upsilon^\mu = -1 \\
	\bar h_{\mu \nu} \hat \upsilon^\nu &= 2  \Phi \tau_\mu \qquad \qquad & \bar h_{\mu \nu} h^{\nu \rho} - \tau_\mu \hat \upsilon^\rho = \delta^\rho_\mu\ ,
\end{align}
%%%%%
and transforming under local U(1) mass transformation as
%%%
\begin{align}
	\delta \bar h_{\mu \nu} &= -\tau_\mu \partial_\nu \sigma - \tau_\nu \partial_\mu \sigma \, , \\
	\delta \hat \upsilon^\mu &= - h^{\mu \nu} \partial_\nu \sigma \, , \\ 
	\delta \Phi &=- \hat \upsilon^\mu \partial_\mu \sigma. 
\end{align}
%%%
The rest of the structures transform only under diffeomorphisms in the usual way. Spacetime measure $e$ can be written in terms of these variables as
%%%
\begin{align}
	e = \sqrt{\frac{\det \bar h}{2 \Phi } }.
\end{align}
%%%
To study the dynamic evolution of systems coupled to a TNC geometry covariantly, it is necessary to introduce a covariant derivative adapted to the generic curved TNC background. There exists no unique metric-compatible connection \cite{Bekaert:2014,Hartong:2015}, so we ask for compatibility with $\{\tau_\mu, h^{\mu \nu} \}$, namely
\begin{align}\label{compatiblity}
	\nabla_\mu \tau_\nu = 0 \ \qquad
    \text{and}
    \qquad \nabla_\mu h^{\rho \sigma} =0 \, , 
\end{align}
%%%
with $\nabla$ the covariant derivative constructed using the affine connection $\Gamma^\rho{}_{\mu \nu}$. A connection satisfying \eqref{compatiblity} and explicitly boost invariant is given by 
%%%
\begin{align}\label{conTNC}
	\Gamma^\rho{}_{\mu \nu} = - \hat \upsilon^\rho \partial_\mu \tau_\nu + \frac{1}{2} h^{\rho \sigma} \left(\partial_\mu \bar h_{\nu \sigma} + \partial_\nu \bar h_{\mu \sigma} - \partial_\sigma \bar h_{\mu \nu} \right)\, .
\end{align} 
%%%
Unlike the Riemannian Christofell connection, $\Gamma^\rho{}_{\mu \nu}$ as defined in \eqref{conTNC} is generically torsionful
%%%
\begin{align}
	 \Gamma^{\lambda}{}_{[\mu \nu]} = - \frac{1}{2} \hat \upsilon^\lambda F_{\mu \nu}\, ,
\end{align}
%%%
where we defined the torsion two form $F_{\mu \nu}$ as 
%%%
\begin{align}
	F_{\mu \nu} = \partial_\mu \tau_\nu - \partial_\nu \tau_\mu .
\end{align}
%%%%
For most cases of physical interest, such torsion vanishes \footnote{This statement is equivalent to saying that for torsionless systems there exists a universal time.}. The connection \eqref{conTNC} is invariant under boosts, however it transforms under U(1) mass transformations as
%%%
\begin{align}
	\delta \Gamma^\rho_{\mu \nu} =  \frac{1}{2}h^{\rho \lambda} \left( F_{\mu \nu} \partial_\lambda \sigma + F_{\lambda \nu} \partial_\mu \sigma + F_{\lambda \nu} \partial_\nu \sigma \right) . 
\end{align} 
%%%
For a vanishing torsion, the connection is invariant under both boost and U(1) transformations. The Riemann tensor $R^\rho{}_{\sigma \mu \nu}$ of the TNC background is defined in the standard way
%%%
\begin{align}
	R^\rho{}_{\sigma \mu \nu} = \partial_\mu \Gamma^\rho_{\nu \sigma} - \partial_\nu \Gamma^\rho_{\mu \sigma} + \Gamma^\rho_{\mu \lambda}\Gamma^\lambda_{\nu \sigma} - \Gamma^\rho_{\nu \lambda}\Gamma^\lambda_{\mu \sigma}. 
\end{align}

\subsection{TNC Submanifolds}

A (p+1)-dimensional TNC submanifold $\Sigma_{p+1}$, with coordinates \footnote{We will use the lower case latin indices $\{a,b,...=0,...,p \}$ to denote the submanifold spacetime indices} $\sigma^a$, embedded in a (d+1)-dimensional TNC manifold $\mathcal{M}_{d+1}$, with coordinates $x^\mu$, is parametrized by the map $X^\mu: \Sigma \rightarrow \mathcal{M}, \mu=0,...,d$ relating the coordinates $\sigma^a$ on $\Sigma_{p+1}$ to the location $X^\mu(\sigma^a)$ on $\mathcal{M}$. In other words, the embedding specifies the location of the surface in the ambient space $\mathcal{M}$. The tangent vectors $u^\mu_a$ to the surface are defined from the embedding map as
%%%%
\begin{align}
	u^\mu_a &= \partial_a X^\mu\, . 
\end{align}
%%%%
The normal 1-forms\footnote{The upper case Latin indices $\{I,J,...= 1,...,d-p \}$ correspond to indices on the transverse directions to the surface. We will consider co-dimension one surfaces, meaning there will only be a single transverse direction, and no upper Latin indices will be used beyond this section.} $n^I_\mu$  will be implicitly defined through the relations
%%%%
\begin{align}\label{normalDef}
n^I_\mu u^\mu_a &= 0 \, , \\
h^{\mu \nu} n^I_\mu n^J_\nu &= \delta^{IJ}, 
\end{align}
%%%%
where $I=1,...,d-p$. Notice that $n^I_\mu$ is not uniquely fixed by \eqref{normalDef} as such relation is invariant under SO(d-p) transformations. 
The inverse tangent vector $u^a_\mu$ and inverse normal 1-form $n^\mu_I$ can be introduced through the orthonormality conditions
%%%%
\begin{align}
	\delta^\mu_\nu &= u^\mu_a u^a_\nu + n^I_\nu n^\mu_I \, , \\
	u^a_\mu n^\mu_I &= 0 \, , \\
	u^\mu_a u^b_\mu &= \delta^b_a \, , \\
	n^\mu_I n^J_\mu &= \delta^J_I.
\end{align}
%%%
In time-like submanifolds the normal vector $n^\mu_I$ satisfies $n^\mu_I \tau_\mu=0$, therefore we can explicitly set $n^{\mu I}=h^{\mu \nu} n^I_\nu$. We can use the tangent vectors and normal 1-forms, as well as their inverses, to project any tensor tangentially or orthogonally to the surface. In particular, the relevant boost-invariant  TNC structure induced on the surface  will be given by the set of objects $\{\tau_a, \bar h_{ab}, \check{\Phi} \}$ and $\{\hat \upsilon^a, h^{ab} \}$ related to the ambient TNC structures through the relations
%%%
\begin{align}
	\tau_a &= u^\mu_a \tau_\mu \, , \\
	\bar h_{ab} &= u^\mu_a u^\nu_b \bar h_{\mu \nu} \, , \\
	\hat \upsilon^a &= u^a_\mu \hat \upsilon^\mu \, , \\
	h^{ab} &= u^a_\mu u^b_\nu h^{\mu \nu} \, , \\
	\check{\Phi} &= \Phi - \frac{1}{2} \hat \upsilon^\mu \hat \upsilon^\nu n^I_\mu n^J_\nu \delta_{IJ} \, , 
\end{align}
%%%
and subject to the orthonormality conditions
%%%
\begin{align}
	\tau_a h^{ab} &= 0, \qquad \qquad & \tau_a \hat \upsilon^a = -1, \\
	\bar h_{ab} \hat \upsilon^b &= 2 \check{\Phi} \tau_a, \qquad \qquad & \bar h_{ab} h^{bc} - \tau_a \hat \upsilon^c = \delta^c_a.
\end{align}
%%%%
This induced TNC structure is invariant under local Milne boosts while transforming under submanifold diffeomorphisms ($\zeta^a$), and U(1) gauge transformations $\sigma$ as 
%%%
\begin{align}
	\delta \tau_a &= \mathcal{L}_\zeta \tau_a \, , \\
	\delta \bar h_{ab} &= \mathcal{L}_\zeta \bar h_{ab} - \tau_a \partial_b \sigma - \tau_b \partial_a \sigma \, , \\ 
	\delta \hat \upsilon^a &= \mathcal{L}_\zeta \upsilon^a - h^{ab}\partial_b \sigma \, , \\
	\delta h^{ab} &= \mathcal{L}_\zeta h^{ab} \, , \\
	\delta \check{ \Phi} &= \mathcal{L}_\zeta \check{ \Phi} - \hat \upsilon^a \partial_a \sigma. 
\end{align}
%%%%
Note that the underlying U(1) transformation is the one associated with the gauge connection $\check{m}_a$, $\delta \check{m}_a=\partial_a \sigma$, related to the bulk one via
%%%%
\begin{align}
    \check{m}_a&= u^\mu_a m_\mu - \frac{1}{2} \upsilon^I \upsilon_I \tau_a. \
\end{align}
%%%%
Covariant differentiation can be defined along the surface direction through the introduction of the covariant derivative $D_a$ satisfying the compatibility conditions
%%%
\begin{align}
	D_a \tau_b &= 0, \qquad \qquad &D_a h^{bc}&=0, \\ 
	D_a \tau_\mu &=0, \qquad \qquad &D_a h^{\mu \nu} &= 0.\ 
\end{align}
%%%
This covariant derivative will act on a mixed index tensor $T^{b \mu}$ as 
%%%
\begin{align}
	D_a T^{b \mu} = \partial_a T^{b \mu} + \gamma^b_{ac} T^{c \mu} + u^\rho_a \Gamma^\mu_{\rho \lambda} T^{b \lambda},
\end{align}
%%%
with the surface affine connection given by 
%%%
\begin{align}
	\gamma^c_{ab} &= -\hat \upsilon^c \partial_a \tau_b + \frac{1}{2} h^{cd}\left(\partial_a \bar h_{bd} + \partial_b \bar h_{ad} - \partial_d \bar h_{ab} \right) \\
	&= \Gamma^c_{ab} + u^c_\mu \partial_a u^\mu_b = u^c_\mu u^\nu_a \nabla_\nu u^\mu_b \, , 
\end{align}
%%%
and associated torsion tensor $F_{ab}$ defined through
%%%
\begin{align}
	2 \gamma^c_{[ab]} = -\hat \upsilon^c F_{ab}.
\end{align}
%%%
A submanifold Riemann tensor $\mathcal{R}^a{}_{b cd}$ can be constructed from $\gamma^a_{ab}$ through the standard form
\begin{align}
\mathcal{R}^a{}_{bcd}= \partial_c \gamma^a_{d b} - \partial_d \gamma^a_{c b} + \gamma^a_{cf} \gamma^f_{d b}- \gamma^a_{df} \gamma^f_{c b}.	
\end{align}
%%%%
The boost invariant extrinsic curvature of the submanifold $K_{ab}{}^I$ can be defined from the covariant derivative of the tangential vector through
%%%
\begin{align}
	K_{ab}{}^I = - u^\mu_a u^\nu_b \nabla_{(\mu} n^I_{\nu)}  =n^I_\mu D_a u^\mu_b + \frac{1}{2} \hat \upsilon^I F_{ab},
\end{align}
%%%
that transforms under mass U(1) transformations as
%%%%
\begin{align}
	\delta K_{ab}{}^I = \frac{1}{2} F_{I a} \partial_b \sigma + \frac{1}{2} F_{Ib} \partial_a \sigma .
\end{align}
%%%%
Notice that the covariant derivative $D$ does not act on normal indices $I$. It is also possible to define a covariant derivative $\mathcal{D}$ that acts on all types of indices. In particular, it is found that 
%%%%
\begin{align}
	\mathcal{D}_a T^I = D_a T^I - \omega_a{}^I{}_J T^J.
\end{align}
%%%%%
The tensor $\omega_a{}^I{}_J$ is called the external rotation tensor and can be interpreted as an SO(d-p) connection. It is defined from the covariant derivative of the normal 1-form
%%%
\begin{align}
	\omega_a{}^I{}_J = n^\mu_J D_a n^I_\mu \, ,
\end{align}
%%%
it is antisymmetric in $I,J$ indices, and transforms under U(1) mass transformations as 
%%%
\begin{align}
	\delta \omega_a{}^I_J = - \frac{1}{2} \left( F_{aJ} \partial^J \sigma + F^I{}_J \partial_a \sigma + \tau^I{}_a \partial_J \sigma \right).
\end{align}
%%%
Notice that for submanifolds with co-dimension one the external rotation tensor vanishes by definition. The curvature associated with the external rotation tensor is known as the outer curvature $\Omega^I{}_{Jab}$ and is given by 
%%%
\begin{align}
	\Omega^I_{J ab} = 2 \partial_{[a} \omega_{b]}{}^I{}_J - 2 \omega_{[a|}{}^I{}_K \omega_{|b]}{}^K{}_J .
\end{align}
%%%%
 The set of submanifold curvatures $\{\mathcal{R}^a{}_{bcd}, K_{ab}{}^I, \Omega^I_{J ab} \}$ are related to the ambient space curvature $R^\rho{}_{\sigma \mu \nu}$ through the integrability conditions known as the Codazzi-Mainardi equation

%%%
\begin{align}\label{Codazzi-Mainardi}
	\mathcal{D}_a \tilde K_{bc}{}^I   - \mathcal{D}_b  \tilde K_{ac}{}^I =  R^I{}_{c ab} + \hat \upsilon^d F_{ab} \tilde K_{dc}{}^I,
\end{align}
%%%
the Gauss-Codazzi equation
%%%
\begin{align}\label{Gauss-Codazzi}
	\mathcal{R}^d_{c ab} = R^d_{cab} + \left( \tilde K_{bc}{}^I \tilde K_{af}{}^J  - \tilde K_{ac}{}^I  \tilde K_{bf}{}^J \right) h^{fd} \delta_{IJ},
\end{align}
and the Ricci-Voss equation
%%%
\begin{align}\label{Ricci-Voss}
	\Omega^I{}_{J ab} = -R^I_{J ab} - 2 h^{cd} \tilde K_{[a|c}{}^I \tilde K_{|b] d}{}^J \delta_{IJ},
\end{align}
%%%%
where in \eqref{Codazzi-Mainardi}-\eqref{Ricci-Voss} we defined
%%%
\begin{align}
	\tilde K_{ab}{}^I \equiv  K_{ab}{}^I - \frac{1}{2} \hat \upsilon^I F_{ab} = n^I_\mu D_a u^\mu_b.
\end{align}
%%%%

\section{Currents and conservation equations}\label{Sec:Currents}

Let's consider the action of a (d-1)-dimensional submanifold $\Sigma$ of a d-dimensional manifold $\mathcal{M}_d$
%%%%
\begin{align}
    S=S_{\Sigma} \left[ \hat \phi, \bar h_{ab},\tau_a, K_{ab} \right],
\end{align}
%%%%%%%%
where $\phi$ are the fields describing the microscopic theory with $\hat \phi$ their value at the sub-manifold. A variation of this action will take the form\footnote{Notice that there is no transverse index on $\mathcal{D}$ or $\delta K$ as the boundary is a co-dimension one surface.}
%%%
\begin{align}\label{deltaS}
	\delta S &=\int_{\partial \mathcal{M}} d^{d-1} \sigma \hat e \left(\left(\mathcal{T}_S\right)^a \delta \tau_a + \frac{1}{2} \left(\mathcal{T}_S\right)^{ab} \delta \bar h_{ab} + \mathcal{D}^{ab} \delta K_{ab} \right), 
\end{align}
%%%%

where $\hat e=\sqrt{\frac{\det \bar h_{ab} }{2 \hat \Phi} }$, $\mathcal{T}^a_S$ is the surface boundary current, $\mathcal{T}^{ab}_S$ is the surface Cauchy stress-mass tensor\cite{Geracie:2016}, and $\mathcal{D}^{ab}$ is the bending moment encoding elastic responses. 
he conservation equation associated with mass conservation can be derived by looking at the U(1) mass transformations 
%%%%
\begin{align}
    \delta \bar h_{ab} &= - \tau_a \partial_b \sigma - \tau_b \partial_a \sigma \, , \\ 
    \delta K_{ab} &= -\frac{1}{2}f_a \partial_b \sigma - \frac{1}{2} f_b \partial_a \sigma \, , 
\end{align}
%%%%
with $f_a= F_{a \lambda} n^\lambda$, and asking for \eqref{deltaS} to be invariant under this set of transformations. This results in the mass conservation equation 
%%%
\begin{align}
    D_a \left[ \left(\mathcal{T}_S\right)^{ab} \tau_b+ \mathcal{D}^{ab} f_b  \right] =0  \, .
\end{align}
  
The remaining equations of motion can be derived by considering Lagrangian variations \cite{Armas:2013,Armas:2020} in \eqref{deltaS}. Lagrangian variations are those that transform the bulk fields but keep the embedding fixed. The Lagrangian variations under diffeomorphisms relevant to \eqref{deltaS} are
%%%
\begin{align}
    \delta \tau_a &= \tau_\rho D_a \xi^\rho - F_{a \rho} \xi^\rho \, , \\
    \delta \bar h_{ab} &=2 \bar h_{\rho (a} D_{b)} \xi^\rho - 2 \xi^\sigma \left[ \bar{\mathcal{K}}_{\sigma(a}\tau_{b)} + \tau_\sigma \tau_{(a} D_{b)}\Phi \right. \\ \nonumber &\left. + \tau_a \tau_b \nabla_\sigma \Phi + F_{\sigma(a} \tau_{b)}\Phi \right] \, , \\
        \delta K_{ab} &= n_\mu D_a D_b \xi^\mu -R^\mu{}_{ba\alpha} \xi^\alpha n_\mu -v K_{ab} F_{\mu \lambda} \xi^\mu n^\lambda  \\ &  + n_\mu  u^\lambda_b D_a\left(\hat \upsilon^\mu F_{\lambda \alpha} \xi^\alpha \right) + \frac{1}{2} v n_\mu F_{ab} n^\lambda \nabla_\lambda \xi^\mu  \, , \nonumber  
\end{align}
%%%%%%%
where $ \Bar{\mathcal{K}}_{\mu \nu}=-\frac{1}{2}\mathcal{L}_{\hat \upsilon}\bar h_{\mu \nu}$ and $v= \hat \upsilon^m n_\mu$. 
Invariance of \eqref{deltaS} under this set of transformations gives the tangential equations of motion
%%%%%
\begin{align}
    D_a \left(\mathcal{T}_S\right)^a  &=\mathcal{K}^\Sigma_{ab} \left( \mathcal{T}_S\right)^{ab}+ 2 \left(\mathcal{T}_S\right)^{ab} \tau_b D_a \check\Phi \\  \nonumber &+  2 \hat \upsilon^d D_a\left( \mathcal{D}^{ab} K_{db} \right) - \hat \upsilon^a \mathcal{D}^{cd} D_a K_{cd} \\
    \nonumber &-2\check{\Phi}D_a \left(\mathcal{D}^{ab} f_b \right)- 2  \mathfrak{a}_c \left( \mathcal{T}_S \right)^c  + 2\check{\Phi} \mathfrak{a}_c \tau_d \left(\mathcal{T}_S \right)^{cd}   \\ \nonumber &+2 \mathfrak{a}_c \hat \upsilon^a K_{ad} \mathcal{D}^{cd}  \, , \\
    D_a \left( \mathcal{T}_S \right)^{a c} &= h^{ca} \mathcal{D}^{f g} D_a K_{f g}- 2 h^{cd} D_a \left( \mathcal{D}^{ab}  K_{d b}  \right)  \\ \nonumber &- \mathfrak{a}_f h^{c b} \bar h_{bd} \left(\mathcal{T}_S \right)^{df} - F_{ab} h^{bc} \left(\mathcal{T}_S \right)^a+\hat \upsilon^c D_a \left( \mathcal{D}^{ab}f_b \right) \\
    \nonumber & -2 F_{db} h^{cd} \hat \upsilon^f K_{fa}\mathcal{D}^{ab}  \, ,
\end{align}
%%%
and the transverse constraint
%%%
\begin{align}\label{YoungLaplace}
    K_{ab} \left(\mathcal{T}_S\right)^{ab} &= D_a D_b \mathcal{D}^{ab} - h^{cd} \mathcal{D}^{ab} K_{ac} K_{bd}  \\  &+ \mathcal{D}^{ab} R_{ab} +v D_a\left(\mathcal{D}^{ab} f_b \right)-v \mathfrak{a}_c \left(\mathcal{T}_S\right)^{cd}\tau_d \nonumber  \\ \nonumber &- f_a \left[ \left(\mathcal{T}_S\right)^a + v D_b \mathcal{D}^{ab}- \hat \upsilon^b K_{bd} \mathcal{D}^{ad} - \frac{v}{2} \mathfrak{a}_b \mathcal{D}^{ab}   \right] \\ \nonumber &- v F_{ab} h^{ac} \left( K_{cd} + \frac{v}{4} F_{cd} \right) \mathcal{D}^{bd} \, ,
\end{align}
%%%%
where we defined $\mathcal{K}^\Sigma_{ab} = - \frac{1}{2} \mathcal{L}_{\hat \upsilon} \bar h_{ab}, R_{ab} = R^\lambda{}_{a \mu b } n^\mu n_\lambda,$ and $\mathfrak{a}_n=\hat \upsilon^m F_{mn}$. This last equation is the shape equation and will determine the embedding function $n_\mu X^\mu$.

\section{Equilibrium Constitutive Relations}\label{Sec:Eq}

Our focus will be on writing out the membrane constitutive relations on equilibrium configurations. Such configurations are characterized by the existence of a time-like killing vector $V^\mu$ that characterizes the time independence of the fluid configuration. In particular, to have an equilibrium configuration, the background should be invariant under the symmetry transformation %triplet $\{V^\mu, \lambda_\mu^V, \Lambda^V \}$, namely 
%%%%%%
\begin{align}
    \mathcal{L}_V \tau_\mu = 0  \qquad \text{and} \qquad
    \mathcal{L}_V \bar h_{\mu \nu} = 2 \tau_{(\mu} \partial_{\nu)} \Lambda^V,
\end{align}
%%%%
where $\Lambda^V$ is the associated gauge parameter. The equivalent statement should hold for the pullback onto the surface $k^a = u^a_\mu k^\mu$, namely
%%%%
\begin{align}\label{eqCond}
    \mathcal{L}_V \tau_a =0 \qquad 
    \text{and} \qquad
    \mathcal{L}_V \bar h_{ab} = 2\tau_{(a} \partial_{b)}\Lambda^K.
\end{align}
%%%%%
An equilibrium state is specified by a given set of gauge invariant physical parameters $\{T, \mu, u^\rho \}$ identified with the temperature, the mass chemical potential, and the four velocity, respectively. In equilibrium, the associated generating function $\mathcal{F}[T,\mu,u]$ should be a function of the background geometry and the symmetry set $\{V^\mu,\Lambda^V\}$. This means that the physical parameters have a solution in terms of the background geometry and the symmetry set; this solution can easily be found to be \cite{Armas:2020,Jain:2020}
%%%%%
%\begin{align}
%    u^\mu = \frac{V^\mu}{V^\mu \tau_\mu} \, , \qquad T = \frac{T_0}{V^\mu \tau_\mu} \ , \qquad \frac{\mu}{T} = \frac{\Lambda^V}{T_0} + \frac{1}{2T} \bar h_{\rho \sigma} u^\rho u^\sigma 
%\end{align}
%%%%%
%and equivalently on the boundary surface
%%%%
\begin{align}\label{eqDefinitions}
    u^a = \frac{V^a}{V^c \tau_c} \, , \qquad T = \frac{T_0}{V^c \tau_c} \, , \\  \nonumber \frac{\mu}{T} = -\frac{1}{2} \frac{\Lambda^V}{T_0} + \frac{1}{2T} \bar h_{c d} u^c u^d,
\end{align}
%%%%
where $T_0$ is a reference temperature, and $u^\mu$ has been normalized to $u^\mu \tau_\mu = 1$. All that has to be done now is to write down the most general generating function $\mathcal{F}[T,\mu,K]$ up to an arbitrary number of derivatives. 
%%%%
The equilibrium conditions \eqref{eqCond} give rise to the conditions
%%%
\begin{align}
    D_a T &=0 \, , \\
    u^a D_a \mu &= 0 \, , \\
    h^{rs} D_s \mu + a^r &=0 \, , \\
    \theta &=0 \, , \\
    \sigma^{a b} &=0 \, ,  
\end{align}
%%%%%%
where we defined $\Delta^a_c = h^{a d} \bar h_{dc}$ and where we used the identifications
%%%%
\begin{align}
     a^d &= u^c D_c u^d  \, , \label{acceleration} \\
    \theta &=\Delta^b_c D_b u^c  \, , \\
    \sigma^{fg} &= \left[ h^{bg} \Delta^f_c + h^{bf}\Delta^g_d - h^{fg} \frac{ \Delta^b_c}{d-2} \right] D_b u^c \, , \\
    \Omega^{fg} &= \frac{1}{2} \left(h^{bg}\Delta^f_c - h^{bf} \Delta^g_c \right) D_b u^c \label{vorticity} \, , 
\end{align}
%%%%

Where $a^d$ is the acceleration, $\theta$ the compresibility, $\sigma^{fg}$ the shear, and $\Omega^{fg}$ the vorticity. To finalize, we can note the following definitions that will come in handy later
%identites and definitions that will come handy later on 
%%%%
\begin{align}
    %u^a u^b K_{ab} = n_\mu a^\mu \, , \\
    \bar u_a &\equiv \bar h_{ab} u^b \, , \\ 
    \bar u^2 &\equiv \bar h_{ab} u^a u^b = \bar u_a u^a \, ,  \\
    \bar a^2 &\equiv \bar h_{cd} a^c a^d = \bar a_d a^d \, , \\
    \bar \Omega^2 & \equiv \Omega^{ab}  \Omega^{cd} \bar h_{ac} \bar h_{bd}.
    %\Omega^a \equiv \Omega^{a \nu} n_\nu \, ,
\end{align}

\subsection{Equilibrium Partition Function}

We need to classify all possible gauge invariant scalars that can be constructed using $\{\mu,T,u^a,K_{ab}, \bar h_{cd}, \tau_d\}$  up to second order in derivatives\footnote{The Canham-Helfrich functional used to derive the biconcave shape of blood cells contains terms of quadratic order in the extrinsic curvature meaning that for consistency we must keep all terms up to the same order in derivatives.}. We are assuming that both $\{T,\mu\}$ are zeroth order while $K_{ab}$ is taken to be first order.  

\begin{table}[hbt!]
\begin{center}
		\begin{tabular}{| c   c        | }
		\hline
		\multicolumn{2}{| l |}{Order zero hydrostatic scalars} \\
		\hline
		$T$ & $\mu$    \\ 
		\hline
		\multicolumn{2}{| l |}{Order one hydrostatic scalars} \\
		\hline
		$F_{(1)}= K_{ab}h^{ab}$ & $F_{(2)} = K_{ab}u^a u^b$ \\ 
		\hline
		\multicolumn{2}{| l |}{Order two hydrostatic scalars} \\
		\hline
			$ S_{(1)}= \left(K_{ab} h^{ab} \right)  \left(K_{cd} h^{cd}\right)$ \qquad  & 
			$S_{(2)}=\left(K_{ab} K_{cd} h^{ac} h^{bd}\right)$ \\ 
            $S_{(3)}=\left(K_{ab} K_{cd} u^a h^{bc} u^d\right)$  \qquad & $ S_{(4)}=\left(K_{ab} u^a u^b\right) \left(K_{cd} u^c u^d\right)$ \\
    $S_{(5)}=a^c a^d \bar h_{cd}$  \qquad   & $S_{(6)}=\left(K_{cd}u^d \right)a^c$ \qquad \\  $S_{(7)}=\Omega^{ab}\Omega^{cd}\bar h_{ac} \bar h_{bd}$   \qquad &   \\
%    &    & & \\
\hline 
	\end{tabular}
\end{center}
\caption{\label{T:allscalarsnoK} 
	Summary of zeroth, first and second order independent inequivalent scalars which may contribute to the equilibrium partition function. Note that the scalars are independent before taking into account the transverse constraint \eqref{YoungLaplace}. The equation of motion \eqref{YoungLaplace} will formally relate the scalars $F_{(1)}$ and $F_{(2)}$ with each other, leaving only one of them as independent. Consequently, $S_{(1)}$ and $S_{(4)}$ will also be related to each order. } 
\end{table}
%%%
%Note that we included some contributions involving the bulk fields through contractions of the normal vector $n_\mu$ and the  fields $\{a^\mu, \Omega^{a\mu}\}$. 
%%%
When constructing table \ref{T:allscalarsnoK} we assumed zero torsion and a flat ambient spacetime. We also ignored terms related to each other through partial integration and used the corresponding integrability condition
%%%
\begin{align}
    \mathcal{R}^d_{cab} &= \left(K_{bc}K_{af} - K_{ac} K_{bf} \right) h^{fd} \, , 
\end{align}
to ignore any explicit curvature terms. Note that the structures in table \ref{T:allscalarsnoK} have already been classified in \cite{Armas:2013,Armas:2013.2} on a relativistic setting. Namely, the scalars shown in table \ref{T:allscalarsnoK} can be regarded as the non-relativistic limit of those already obtained in \cite{Armas:2013,Armas:2013.2}. The equilibrium partition function can then be written as
%%%
\begin{align}\label{partitionFs}
    \mathcal{F}= \int d^{d-1} \sigma \hat e &\left[ \chi(T,\mu) + \sum_{i=1}^2 a^{(1)}_i(T,\mu) F_{(i)} + \right. \\ &\left. + \sum_{i=1}^{7} a^{(2)}_i (T,\mu) S_{(i)} \right] \nonumber.
\end{align}
%%%%
Where $\chi(T,\mu)$ is related to the surface tension \footnote{See equation \eqref{chiSurfaceRel} for the relation between $\chi$ and the surface tension and the spontaneous curvature of a biophysical membrane.}, and $\{a^{(1)}_i(T,\mu),a^{(2)}_i(T,\mu)\}$ are generic functions of the temperature and chemical potential. From this point onward, and for the sake of simplicity, we will omit the functional dependence of these aforementioned functions. Before moving forward, it is instructive to note that \eqref{partitionFs} can be written as
%%%
\begin{align}\label{PartitionExplicit}
    \mathcal{F}& = \int d^{d-1} \sigma \hat e \left[ \chi + a^{(1)}_1 K + a^{(1)}_2 \left( a \cdot n \right) + a^{(2)}_1 K^2 \right. \\
    \nonumber &a^{(2)}_2 \left(K \cdot K \right) + a^{(2)}_3 \left( n \cdot \Omega \right)^2 + a^{(2)}_4 \left(a \cdot n \right)^2 + a^{(2)}_5 \bar a^2   \\
    & \left.\vphantom{\int} + a^{(2)}_6 \left( \bar a \cdot \left(n \cdot \Omega\right) \right) + a^{(2)}_7 \bar \Omega^2 \right] , \nonumber
\end{align}
%%%%
where we used the following shorthand notation for the spatial traces of the extrinsic curvature
%%%%
\begin{align}
    K=h^{cd}K_{cd}  , \qquad \left(K \cdot K \right)= h^{ac}h^{bd} K_{ab} K_{cd}  ,
\end{align}
%%%
%%%%
as well as the following shorthand notation for the normal components of acceleration and vorticity \footnote{The normal components of acceleration and vorticity show up due to the identities $u^\mu n_\mu=0$ and $K_{ab}u^b = n_\mu D_a u^\mu $. In equilibrium it holds that $h^{ac}K_{ab}u^b = n_\rho u^c_\sigma \Omega^{\rho \sigma} $ and $K_{ab}u^a u^b = n_\mu u^\nu \nabla_\nu u^\mu$  }  
%%%
\begin{align}
    n_\nu u^\mu \nabla_\mu u^\nu = n_\nu a^\nu &  \equiv \left(a \cdot n \right) ,  \\
    n_\rho \Omega^{\rho \beta}  & \equiv \left(n \cdot \Omega \right)^\beta ,   
\end{align}
%%%%
%%%
and the following shorthand notation for some relevant contractions
%%%
\begin{align}
    \left(n \cdot \Omega \right)^\alpha \bar h_{\alpha \beta} \left( n \cdot \Omega \right)^\beta  & \equiv  \left(n \cdot \Omega\right)^2 , \\
    \left(n \cdot \Omega \right)^\alpha \bar h_{\alpha \beta} a^\beta& \equiv \left( \bar a \cdot \left(n \cdot \Omega\right) \right) .
\end{align}
From \eqref{PartitionExplicit} we can note that the coefficients $\{ a^{(1)}_1, a^{(2)}_1, a^{(2)}_2 \}$ will be associated with the contributions of the extrinsic and intrinsic curvatures to the equilibrium partition function. Coefficients $\{ a^{(1)}_2,a^{(2)}_4 \}$ will be associated with the contributions of the normal acceleration to the equilibrium partition function.
The coefficient $a^{(2)}_3$ will be associated with the contribution of the normal vorticity to the equilibrium partition function.
The coefficient $a^{(2)}_5$ will be associated with the contribution of the tangential acceleration to the equilibrium partition function.
The coefficient $a^{(2)}_7$ will be associated with the contribution of the tangential vorticity to the equilibrium partition function.
Finally, $a^{(2)}_6$ will be associated with the contribution of the spatial inner product between the normal vorticity and the tangential acceleration. After taking into account this form for $\mathcal{F}$, we note that the variation of the equilibrium partition function will take the form
%%%%
    \begin{align}
    \delta \mathcal{F} &=  \int d^{d-1} \sigma \hat e \left(\frac{h^{ab} \delta \bar h_{ab} - 2 \hat \upsilon^a \delta \tau_a }{2} \right)  \left[ \chi + \sum_{i=1}^2 a^{(1)}_i F_{(i)} + \sum_{i=1}^{7} a_i^{(2)} S_{(i)} \right] \\ &+  \int d^{d-1} \sigma \hat e \left[ s \delta T + n \delta \mu  + \sum_{i=1}^2 \left(s^{(1)}_i \delta T + n^{(1)}_i \delta \mu \right) F_{(i)} \right. \nonumber \\
    & \left. + \sum_{i=1}^{7} \left( s_{i}^{(2)}\delta T + n_i^{(2)} \delta \mu \right)S_{(i)}    \right] \nonumber \\
     & \nonumber  + \int d^{d-1} \sigma \hat e \left[ \sum_{i=1}^{2}a^{(1)}_i \delta F_{(i)} + \sum_{i=1}^{7} a^{(2)}_i \delta S_{(i)} \right] , 
\end{align}
%%%%
where we used the identity $\delta \hat e =  \frac{\hat e}{2}\left( h^{ab}\delta \bar h_{ab} - 2 \upsilon^a \delta \tau_a  \right)$, and defined the surface entropy density $s$, the surface particle density $n$, and the surface susceptibilities $\{ s^{(1)}_i,s^{(2)}_i,n^{(1)}_i,n^{(2)}_i \}$ as
%%%

\begin{align}\label{equationsOfState}
    s &= \left( \frac{\partial \chi}{\partial T} \right)_\mu \, , \qquad &n &= \left( \frac{\partial \chi}{\partial \mu} \right)_T  \, , \\ 
    s^{(1)}_i &= \left(\frac{\partial a^{(1)}_i}{\partial T} \right)_\mu \, , \qquad &n^{(1)}_i &= \left( \frac{\partial a^{(1)}_i}{\partial \mu} \right)_T \, , \nonumber \\
    s^{(2)}_i &= \left(\frac{\partial a^{(2)}_i}{\partial T} \right)_\mu \, , \qquad &n^{(2)}_i &= \left( \frac{\partial a^{(2)}_i}{\partial \mu} \right)_T .\nonumber  
\end{align}

%%%%%
%%%%%
To find the relevant currents we will  use the following variations of \eqref{eqDefinitions}
%%%%%
\begin{align}
    \delta T &= -  T u^a \delta \tau_a \, , \\
    \delta \mu &= -\left( \mu + \frac{1}{2} \bar u^2  \right)u^a \delta \tau_a + \frac{1}{2} u^a u^b \delta h_{ab} \, , \\
    \delta u^a &= - u^a u^b \delta \tau_b \, , 
\end{align}
%%%%
the variation of \eqref{acceleration} given by
%%%
\begin{align}
    \delta a^d &= \delta u^c D_c u^d + u^c \left( D_c \delta u^d + \delta \Gamma^{d}_{cf} u^f \right) \\ \nonumber &-\delta h^{cd} D_c \mu - h^{cd} D_c \delta \mu  \, ,
\end{align}
%%%%%
also the variation of \eqref{vorticity}
%%%
\begin{align}
    \delta \Omega^{fg} \Omega^{rs} \bar h_{fr} \bar h_{gs} &= \frac{1}{2} \left[ \Omega^{af} \Omega^{bg} \bar h_{ab} \bar h_{fg} \right]u^a \delta \tau_a  \\ \nonumber &+\frac{1}{2}\left[\bar u_a \bar h_{b c} \Omega^{ab} -D_c \left( \mu + \frac{1}{2}\bar u^2 \right)   \right] \Omega^{cd} \delta \tau_d    \\ \nonumber & - \frac{1}{2} \delta \bar h_{cd} \Omega^{cf} \Omega^{dg} \bar h_{fg} - \frac{1}{2} \bar h_{bc} \Omega^{cf} u^b u^d D_f \delta \tau_d \\ \nonumber  &- \frac{1}{2} \Omega^{cd} u^b D_c \delta h_{bd}   \, ,
\end{align}
as well as the variation of the connection 
%%%%
\begin{align}
    \delta \Gamma^a_{cd} &=- \hat \upsilon^a D_c \delta \tau_d + \frac{1}{2}  \delta h^{a b} \left(2D_{(c} \bar h_{d)b} - D_b \bar h_{c d} \right) \\ \nonumber &+ \frac{1}{2} h^{ab}   \left(D_{(c}  \delta\bar h_{d)b} - D_b  \delta\bar h_{c d} \right) ,
\end{align}
%%%%
and the geometric identities 
%%%%
\begin{align}
    \delta h^{ab} &= - h^{ac} h^{bd} \delta \bar h_{cd} + 2 \hat \upsilon^{(a} h^{b)c} \delta \tau_c \, , \\
    \delta \upsilon^a &= \left(2 \check{\Phi} h^{ab}  + \hat \upsilon^a \hat \upsilon^b \right) \delta \tau_b- h^{ac} \hat \upsilon^d \delta \bar h_{cd} .
\end{align}
%%%%
    Using \eqref{deltaS}, the surface energy current is found to be
%%%%
\begin{align}\label{surfaceEnergyEq}
\left(\mathcal{T}_\mathcal{S}\right)^c &=-\chi \hat \upsilon^c -\left( \epsilon + \chi + \frac{n}{2}\bar u^2 \right)u^c - \left( \sum_{i=1}^2 \zeta^{(1) c}_i F_{(i)} + \sum_{i=1}^{7} \zeta^{(2) c}_i S_{(i)} \right) \\ &+ \left[ 2 \left(a_1^{(1)}+ 2 a_1^{(2)} F_{(1)} \right) \hat \upsilon^a K_{ab} + 4 a_2^{(2)} \hat \upsilon^a \left(K_{ar} h^{rs} K_{sb}\right) \right. \nonumber \\
& \left. + 2 a_3^{(2)}\hat \upsilon^a \left(K_{ar}u^r\right)\left( K_{db}  u^d\right)  \right] h^{bc} \nonumber \\ 
\nonumber &- \left[ 4 a^{(2)}_5 \bar  a_f D_g \mu  + 2a^{(2)}_{6} K_{af} u^a D_g \mu \right] \upsilon^{(f}h^{g) c} \nonumber \\
&- \left[2 a^{(2)}_5 h^{fg} D_f D_g \mu +  h^{bf} D_f   \left( a^{(2)}_6 K_{ab}u^a \right) \right] \left(\mu + \frac{1}{2} \bar u^2 \right) u^c  \nonumber \\ 
\nonumber & + \left[2 a^{(2)}_{7} \bar u_a \bar h_{bd}\Omega^{ab} -2 a^{(2)}_{7}  D_d \left(\mu + \frac{1}{2}\bar u^2\right)  \right]\Omega^{cd}\nonumber \\
&+ \left[ D_f \left(a^{(2)}_{7} \bar h_{bc} \Omega^{cf} u^b \right)-a^{(2)}_{7} \Omega^{af} \Omega^{bg}\bar h_{ab} \bar h_{fg}\right]u^c \nonumber \\
&+ \left[2 \left(a_2^{(1)} + 2 a_4^{(2)} F_{(2)} \right)K_{ab} u^a u^b - 2 a_3^{(2)} \left(K_{ar} h^{rs} K_{sb} \right)u^a u^b \right. \nonumber \\
& \left. + a^{(2)}_{6} K_{fg} a^f u^g\right] u^c, \nonumber
\end{align}
%%%%
where the tensors $\{\zeta^{(1)c}_i, \zeta^{(2)c}_i \}$ are defined as 
\begin{align}
    \zeta^{(1)c}_i &= a^{(1)}_i \hat \upsilon^c + \left[s^{(1)}_i T + n^{(1)}_i \left(\mu + \frac{1}{2} \bar u^2 \right)  \right] u^c \, , \\
    \zeta^{(2)c}_i &= a^{(2)}_i \hat \upsilon^c + \left[s^{(2)}_i T + n^{(2)}_i \left(\mu + \frac{1}{2} \bar u^2 \right)  \right] u^c \, , 
\end{align}
and where we have defined the internal energy $\epsilon$ as \cite{Armas:2020}
%%%
\begin{align}\label{energyRel}
    \epsilon = - \chi + T s + n \mu ,
\end{align}
%%%
Note that using \eqref{equationsOfState}, the relation \eqref{energyRel} is equivalent to the thermodynamics first law 
%%%
\begin{align}
    d\epsilon &=  T ds + \mu dn   .
\end{align}

Using \eqref{deltaS}, the surface Cauchy stress-mass tensor is found to be
%%%%
\begin{align}\label{CauchuEq}
    \left(\mathcal{T}_{\mathcal{S}}\right)^{cd} &= \chi h^{cd} + n u^c u^d + \left( \sum_{i=1}^2 \xi^{(1)cd}_i F_{(i)} + \sum_{i=1}^7 \xi^{(2)cd}_i S_{(i)} \right) \\
    & - \left[2 \left(a_1^{(1)} + 2 a_1^{(2)} F_{(1)}\right) K_{ab}  +4 a_2^{(2)} \left(K_{ar} h^{rs} K_{sb}\right) + 2 a_3^{(2)} \left( K_{ar} u^r\right)\left( K_{bs} u^s \right) \right] h^{a(c} h^{d)b} \nonumber \\
    \nonumber &+ \left[4 a^{(2)}_5 \bar a_f  D_g \mu + 2 a^{(2)}_6 K_{af} u^a D_g \mu \right]h^{cf} h^{dg}  + \left[ a^{(2)}_5 2  h^{fg}D_f \bar a_g + a^{(2)}_6 h^{bf} D_f\left( K_{ab}u^a \right) \right] u^c u^d \\
    \nonumber & +\left[2 a^{(2)}_7 \Omega^{ r a} \Omega^{ s b} \bar h_{rf} \bar h_{s g} \bar h_{ab} \right]h^{f(c} h^{d)g} - 2 D_b\left(a^{(2)}_7 \Omega^{b(c}u^{d)} \right) , 
\end{align}
%%%%

where the tensors $\{\xi^{(1)cd}_i, \xi^{(2) cd}_i\}$ are defined as 
\begin{align}
    \xi^{(1)cd}_i &= a^{(1)}_i h^{cd}+n^{(1)}_i u^c u^d \, , \\
    \xi^{(2)cd}_i &= a^{(2)}_i h^{cd} + n^{(2)}_i u^c u^d \, . 
\end{align}

Finally, using \eqref{deltaS}, the bending moment is found to be
%%%%
\begin{align}\label{bendingMomentEq}
    \mathcal{D}^{cd} &= a_1^{(1)} h^{cd} + a_2^{(1)} u^c u^d + \mathcal{Y}^{cd ab} K_{ab} + a^{(2)}_6 a^c u^d ,
\end{align}
%%%
with $\mathcal{Y}^{cd ab}$ the Young modulus
%%%
\begin{align}
    \mathcal{Y}^{cdab} &= 2 a_1^{(2)} h^{cd} h^{ab} + 2 a_2^{(2)} h^{ac} h^{bd} \\ \nonumber &+ 2 a_3^{(2)} u^{(c} h^{d)b} u^a + 2 a_4^{(2)} u^c u^d u^a u^b .
\end{align}
%%%%
The explicit form of the variations that were used to obtain \eqref{surfaceEnergyEq},\eqref{CauchuEq}, and \eqref{bendingMomentEq} are shown in appendix A. In this appendix, the individual contributions of each term to the currents is also shown. 

\subsection{Axisymmetric configurations in 3+1 flat ambient spacetime  }

For applications concerning biological systems, such as the lipid bilayer membranes discussed in \cite{Helfrich,Armas:2020}. It is enough to restrict the ambient spacetime geometry to 
%%%%
\begin{align}
    \tau_\mu &= \delta_\mu^t \, , \qquad &h_{\mu \nu} &= \delta^i_\mu \delta_\nu^j \delta_{ij} \, ,  \\ \nonumber \hat \upsilon^\rho &= - \delta^\rho_t \, , &h^{\mu \nu} &= \delta^\mu_i \delta^\nu_j \delta^{ij} \, , \quad &m_\mu&=0,
\end{align}
%%%
where we assume a time direction $t$ and a Cartesian three-dimensional space with coordinates $x^i=(x,y,z)$. The equilibrium condition on this background, with gauge parameter $\Lambda^K=0$, has a solution for the Killing vector $V^\mu$ given by 
%%%
\begin{align}
   % V= c_{0} \partial_t +  \left(c_3y- c_2 z  \right)\partial_x + \left(c_1 z -c_3 x \right)\partial_y + \left(c_2 x -  c_1 y \right) \partial_z, \\
        V&= c_{0} \partial_t +  c_1 \left(z \partial_y - y \partial_z \right)+ c_2 \left(x \partial_z - z \partial_x \right) \\ \nonumber &+ c_3 \left(y \partial_x - x \partial_y \right),
\end{align}
%%%
where $\{c_0,c_1,c_2,c_3\}$ are constants that can be chosen at our convenience. We will be interested in studying membranes subjected to axisymmetric flows, therefore, we need to set $c_1=c_2=0$. This corresponds to a flow on the ambient spacetime with velocity 
\begin{align}
    u^\mu &= \delta^\mu_t +  \frac{\omega}{2} \left(x \delta^\mu_y -y \delta^\mu_x \right) ,
\end{align}
%%%
where we have defined $\omega=-\frac{2c_3}{c_0}$. Notice that this corresponds to a flow with a constant spatial vorticity vector $\boldsymbol{\omega}=(0,0,\omega)$. Membranes with axisymmetric symmetry can be parametrized through the embedding map\cite{Armas:2020}:
%%%
\begin{align}\label{embedding}
    X^\mu = \begin{pmatrix}
        t \\ \rho \cos 
        \varphi 
        \\ \rho \sin 
        \varphi
        \\ z_0 + \int_0^\rho d\tilde \rho \tan \psi(\tilde \rho) 
    \end{pmatrix} .
\end{align}
%%%
The coordinates on the membrane will then be $(t,\rho,
\varphi
)$. Under this embedding, it is easy to find that
\begin{align}
    \tau_a &= \delta^t_a \, , \quad &\bar h_{ab} &= \left( \sec^2 \psi \right) \delta^\rho_a \delta^\rho_b  + \rho ^2  \delta^
    \varphi
    _a \delta^
    \varphi
    _b \, , \\ \nonumber \hat \upsilon^a &= - \delta^a_t \, , \quad &h^{ab}&= \left(\cos^2 \psi \right) \delta^a_\rho \delta^b_\rho + \frac{\delta^a_
    \varphi 
    \delta^b_
    \varphi
    }{\rho^2}  .
\end{align}
%%%
The associated fluid velocity $u^a$ will be given by 
\begin{align}
    u^a = \delta^a_t + \frac{\omega}{2} \delta^a_
    \varphi .
\end{align}
%%%
%Where $u_\phi$ is a constant given by $u_\phi=-\frac{c_3}{V_t}$. 
The acceleration, compressibility, shear, and vorticity will be given by 
\begin{align}
    a^b &=\left( - \frac{\omega^2}{4} \rho \cos^2 \psi \right) \delta^b_\rho  \, , \\  
    \theta &= 0 \, ,\\ 
    \sigma^{ab} &= 0 \, , \\ 
    \Omega^{ab} &=\left(  \frac{\omega}{2 \rho} \cos^2 \psi \right) \left(\delta^a_\rho \delta^b_
    \varphi 
    -\delta^a_
    \varphi 
    \delta^b_\rho\right).
\end{align}
%%%
The equilibrium condition implies that temperature is constant and given by $T=\frac{T_0}{c_0}$, while the chemical potential $\mu$ will be a function of $\rho$ given by 
%%%
\begin{align}\label{chemicalPotential}
  \mu(\rho) = \frac{\omega^2 \rho^2 }{8} .
\end{align}
%%%%
Notice that the chemical potential \eqref{chemicalPotential} is equal to the rotational kinetic energy per unit mass of a fluid element rotating with an angular velocity $\frac{\omega}{2}$ at a distance $\rho$ from the rotation axis.
The extrinsic curvature in this configuration is given by 
%%%
\begin{align}
    K_{ab}=  -\left(\psi' \sec \psi \right) \delta^\rho_a \delta^\rho_b - \left( \sin \psi \right) \delta_a^
    \varphi\delta_b^\varphi.
\end{align}
%%%
Using all this information, we show in table \ref{T:allscalars} all the scalars from table \ref{T:allscalarsnoK} in terms of the embedding variables.  
Note that the normal and tangential acceleration contributions are consistent with the centripetal acceleration of a fluid element, and note that the normal vorticity is orthogonal to the tangential acceleration. Note that for any function $f(T,\mu)$, it will be true that in equilibrium
%%%
\begin{align}
   D_b f &= \left( \frac{\partial f}{\partial \mu}\right) D_b \mu , \\
   D_a D_b f &= \left( \frac{\partial^2 f}{\partial \mu^2} \right) \left( D_a \mu \right) \left( D_b \mu \right) + \left(\frac{\partial f} {\partial \mu}\right) D_a D_b \mu .  
\end{align} 

\begin{table}[hbt!]
\begin{center}
		\begin{tabular}{| c     | }
		\hline
		Order zero hydrostatic scalars \\
		\hline
		$T$ \\ $\mu= \frac{\omega^2 \rho^2}{8}$   \\ 
		\hline
		Order one hydrostatic scalars \\
		\hline
		$F_{(1)}= K_{ab}h^{ab} = - \frac{\sin\psi}{\rho} - \psi' \cos\psi $ \\ $F_{(2)} = K_{ab}u^a u^b = - \frac{1}{4} \omega^2 \rho \sin \psi $  \\ 
		\hline
		Order two hydrostatic scalars \\
		\hline
			$ S_{(1)}= \left(K_{ab} h^{ab} \right)  \left(K_{cd} h^{cd}\right)= \frac{\left(\psi' \cos \psi + \sin \psi \right)^2}{\rho^2} $ \\
			$S_{(2)}=\left(K_{ab} K_{cd} h^{ac} h^{bd}\right)= \frac{\sin^2 \psi}{\rho^2}+ \left(\psi'\right)^2 \cos^2 \psi$ \\ 
            $S_{(3)}=\left(K_{ab} K_{cd} u^a h^{bc} u^d\right) = \frac{1}{4} \omega^2 \sin^2 \psi$  \\ 
            $ S_{(4)}=\left(K_{ab} u^a u^b\right) \left(K_{cd} u^c u^d\right) = \frac{1}{16} \omega^4 \rho^2 \sin^2 \psi$ \\
    $S_{(5)}=a^c a^d \bar h_{cd}=\frac{1}{16} \omega^4 \rho^2 \cos^2 \psi$ \\ $S_{(6)}=\left(K_{cd}u^d \right)a^c=0$ \\ $S_{(7)}=\Omega^{ab}\Omega^{cd}\bar h_{ac} \bar h_{bd} = \frac{1}{2}\omega^2 \cos^2 \psi$     \\
%    &    & & \\
\hline 
	\end{tabular}
	\caption{\label{T:allscalars} 
	Summary of zeroth, first and second order independent inequivalent scalars for an axisymmetric system.}
\end{center}
\end{table}
%%%

\section{Young-Laplace equation for  axisymmetric solutions for Canham-Helfrich-like membranes} \label{Sec:YL}

The equation describing the shape of the membrane is given by \eqref{YoungLaplace} and can be considered a generalization of the Young-Laplace equation. In flat ambient spacetime, it takes the form
%%%
\begin{align}
  D_a D_b \mathcal{D}^{ab} -  \left( \mathcal{T_S} \right)^{ab} K_{ab}  - h^{cd} \mathcal{D}^{ab} K_{ac} K_{bd}  =0.
\end{align}
%%%
 In this section, we will focus on the immediate generalization of the case studied in \cite{Armas:2020}. This generalization will consist of keeping only the terms in the equilibrium partition function that depend only on the vorticity through the chemical potential; that is, we will keep only the elastic contributions. This means keeping $\chi, a^{(1)}_1, a^{(2)}_1,$ and $a^{(2)}_2$ as arbitrary functions of the temperature and chemical potential while setting all other $a^{I}_{i}$ coefficients to zero\footnote{The explicit form of the full Young-Laplace equation taking all coefficients into account is quite cumbersome and by itself it is not very instructive, so we will leave its full analysis for future work.}. Therefore, the equilibrium partition function will simply look like 
%%%
\begin{align}\label{partition1}
    \mathcal{F}= \int d^{3}\sigma e\left[ \chi + a_1^{(1)} K +a_1^{(2)} K^2 + a_2^{(2)} \left(K \cdot K\right)  \right],
\end{align}
%%%
The relevant constitutive relations will take the form 
%%%
\begin{align}
    \left(\mathcal{T_S} \right)^{cd} &= \left[\chi+a_1^{(1)} K + a_1^{(2)} K^2 +a_2^{(2)} \left(K \cdot K \right)  \right] h^{cd} \\\nonumber &+ \left[n + n^{(1)}_1 K+n^{(2)}_1 K^2 +n^{(2)}_2 \left(K \cdot K \right) \right] u^c u^d \\ \nonumber &- 2 \left( a^{(1)}_1 + 2 a^{(2)}_1 K \right) h^{ac}h^{bd} K_{ab}\\ \nonumber &- 4 a^{(2)}_2 h^{rs}h^{ac}h^{bd} K_{ar} K_{sb}, \\ 
    \mathcal{D}^{cd} &= a_1^{(1)}h^{cd}+ 2 a_1^{(2)}  h^{cd} K + 2 a_2^{(2)}h^{ac}h^{bd} K_{ab}.
\end{align}
%%%
We can first compute the parts that appear in the Young-Laplace equation without covariant derivatives:
%%%
\begin{align}\label{termKT}
    \left( \mathcal{T}_S \right)^{ab} K_{ab} &= \chi K + a^{(1)}_1 K^2 + a_1^{(2)} K^3  \\ \nonumber & + a_2^{(2)}K \left(K \cdot K \right) + n \left(a \cdot n \right) + n_1^{(1)} K \left(a \cdot n \right) \\ \nonumber &+ n_1^{(2)} K^2 \left(a \cdot n \right)+ n^{(2)}_2 \left(K \cdot K \right) \left(a \cdot n \right) \\ \nonumber  &- 2 a_1^{(1)} \left(K \cdot K \right) - 4 a_1^{(2)} K \left(K \cdot K \right)- 4 a_2^{(2)} K_3  ,
\end{align}
%%%
\begin{align}\label{termKT2}
h^{cd}\mathcal{D}^{ab}K_{ac}K_{bd} &= a_1^{(1)} \left(K \cdot K \right) + 2 a_1^{(2)} K \left(K \cdot K \right) \\ \nonumber & + 2 a_2^{(2)} K_{ab} K_{cd} K_{rs} h^{bc} h^{dr} h^{sa}, 
\end{align}
%%%
where we have used that $K_{cd}u^c u^d=\left(a \cdot n \right)$.
%%%
Combining \eqref{termKT} and \eqref{termKT2}, we end up with
%%%

    \begin{align}\label{ThDkkEq}
     \left(\mathcal{T_S} \right)^{cd}+ h^{cd}\mathcal{D}^{ab}K_{ac}K_{bd} &= \chi K + a_1^{(1)}K^2 \\ \nonumber &  - a_1^{(1)}\left(K \cdot K \right) \\
     \nonumber &+ \left[a_1^{(2)}+a_2^{(2)} \right]\left[K^2- 2 \left(K \cdot K \right) \right] K  \\ \nonumber &+n \left(a \cdot n \right) + n_1^{(1)} K \left(a \cdot n \right) \\ \nonumber & + n_1^{(2)}K^2 \left(a \cdot n \right) \\ \nonumber & + n_2^{(2)} \left(K \cdot K \right) \left(a \cdot n \right),
\end{align}
%%%
where we used the following identity between traces
%%%
\begin{align}
  K_{ab} h^{bc} K_{cd} h^{dr} K_{rs} h^{sa} = \frac{3}{2} K \left(K \cdot K \right)- \frac{1}{2}K^3 .
\end{align}
%%%
We can now take a look at the term with derivatives of the bending moment
%%%

\begin{align}\label{ddDeq}
    D_{c}D_{d} \mathcal{D}^{cd} &= h^{cd} D_c D_d a_1^{(1)} + 2 h^{cd} \left[  K D_c D_d a_1^{(2)} + 2\left(D_c K \right) \left(D_d a_1^{(2)} \right) \right. \\
    & \left. + a_1^{(2)}D_c D_d K  \right] \nonumber \\ \nonumber & + 2 h^{ac} h^{bd} \left[K_{ab} D_c D_d a^{(2)}_2 + 2 \left(D_c K_{ab} \right) \left(D_d a^{(2)}_2 \right)+ a^{(2)}_2 D_c D_d K_{ab}    \right] \\
    &=\nonumber  2\left(a^{(2)}_1+  a_2^{(2)}\right) D^2 K  +  \left( \beta^{(1)}_1 +2 \beta^{(2)}_1 K  \right)\bar a^2 \nonumber \\
    &- \left( n^{(1)}_1 + 2 n^{(2)}_1 K \right) \left(D \cdot a \right) \nonumber  \\
    \nonumber &+ 2 \beta^{(2)}_2 \left( K_{cd} a^c a^d \right) -  2 n^{(2)}_2 \left( K_{cd} h^{ac}  D_a a^d \right) - 4 \left( n^{(2)}_1 + n^{(2)}_2 \right) a^c D_c K,       
\end{align}
where we used the shorthand notation $D^2=h^{cd}D_c D_d$, $D\cdot a = D_c a^c$, the Codazzi-Mainardi equation, and where we defined
%%%
\begin{align}
   \beta^{(1)}_{i} & = \frac{\partial n^{(1)}_i}{\partial \mu} \, , \qquad \beta^{(2)}_i = \frac{\partial n^{(2)}_i}{\partial \mu}.
\end{align}
%%%
Combining \eqref{ThDkkEq} and \eqref{ddDeq}, we find that the Young-Laplace equation takes the form 
%%%
\begin{align}\label{YoungLaplacePre}
    2\left(a^{(2)}_1+  a_2^{(2)}\right) D^2 K  +  \left( \beta^{(1)}_1 +2 \beta^{(2)}_1 K  \right)\bar a^2 - \left( n^{(1)}_1 + 2 n^{(2)}_1 K \right) \left(D \cdot a \right)   \\
    \nonumber + 2 \beta^{(2)}_2 \left( K_{cd} a^c a^d \right) -  2 n^{(2)}_2 \left( K_{cd} h^{ac}  D_a a^d \right) - 4 \left( n^{(2)}_1 + n^{(2)}_2 \right) a^c D_c K \nonumber \\ \nonumber 
  -\chi K - a_1^{(1)}K^2  + a_1^{(1)}\left(K \cdot K \right) - \left(a_1^{(2)}+a_2^{(2)} \right)\left(K^2- 2 \left(K \cdot K \right) \right) K \\ -n \left(a \cdot n \right) - n_1^{(1)} K \left(a \cdot n \right) - n_1^{(2)}K^2 \left(a \cdot n \right) - n_2^{(2)} \left(K \cdot K \right) \left(a \cdot n \right) &= 0 .\nonumber 
\end{align}
%%%%
To understand the possible applications and interpretations of the solutions of \eqref{YoungLaplacePre}, we can recall that, in the context of biophysics, one possibility is to interpret the partition function \eqref{partition1} as the energy functional of a biophysical membrane \cite{Armas:2020}. In particular, we can relate the coefficients $\{ \chi, a^{(1)}_1, a^{(2)}_1, a^{(2)}_2\}$ to those appearing on the Canham–Helfrich free energy model for lipid vesicles. The Canham–Helfrich free energy has the typical form \cite{Hiroyoshi1996}
%%%
\begin{align}\label{CH-action}
   \mathcal{F}_{CH} = \int d^3 x \left[\gamma + \frac{\kappa}{2} \left(K-K_0 \right)^2 + \bar \kappa \mathcal{R}   \right],
\end{align}
%%%%
 where $\mathcal{R}= h^{ac} \mathcal{R}^d{}_{adc}$ is the spatial Ricci scalar. The coefficients that appear in \eqref{CH-action} are the surface tension $\gamma$, the bending modulus $\kappa$, the spontaneous curvature $K_0$, and the Gaussian modulus $\bar \kappa$. The partition function \eqref{partition1} can be related to the free energy functional \eqref{CH-action} through the identifications 
%%%%
\begin{align}
    \chi & = \gamma + \frac{\kappa K_0^2}{2} ,\label{chiSurfaceRel} \\
    a^{(1)}_1  &=-\kappa K_0 ,  \\
    a^{(2)}_1& = \frac{\kappa}{2} + \bar \kappa , \\
    a^{(2)}_2 & =-\bar \kappa ,
\end{align}
%%%%
where we used the identity $\mathcal{R}=K^2 - (K \cdot K)$. Using the Canham-Helfrich coefficients, the Young-Laplace equation \eqref{YoungLaplacePre} takes the form 

   %%%
\begin{align}\label{YoungCoefChange}
   &-K \gamma + \kappa \left[  D^2 K - \frac{1}{2}\left(K - K_0 \right)^2 K   + \left(K-K_0 \right) \left(K \cdot K \right)\right]- n_\gamma \left(a \cdot n\right) \nonumber \\
   &+ \kappa n_0 \left[ \left(D \cdot a \right) + \left(a \cdot n \right)\left( K-K_0 \right)  \right] - 2 n_0 n_\kappa \bar a^2 \nonumber  \\ 
   & - n_\kappa \left[2a^m D_mK+ \frac{1}{2} \left(K-K_0 \right)^2 \left(a \cdot n \right) + \left(D \cdot a \right) \left(K-K_0\right) \right]  - \beta_0 \kappa \bar a^2  \nonumber \\ 
   & + \beta_{\kappa}\left(K-K_0 \right)\bar a^2+  n_{\bar \kappa} \left[2 K_{cr}h^{rs}D_s a^c  -2K \left(D \cdot a \right) + \left(a \cdot n \right) \left( (K\cdot K)- K^2 \right) \right] \nonumber \\
   & + 2\beta_{\bar \kappa} \left( K \bar a^2 - K_{cd}a^c a^d\right)  = \Delta p , 
\end{align}
%%%% 
%%%%
where the coefficients $\{n_\gamma, n_0, n_\kappa, n_{\bar \kappa}, \beta_0,\beta_{\kappa},\beta_{\bar \kappa} \}$ are defined as
%%%%%%
\begin{align}
    n_0&=\frac{\partial K_0}{\partial \mu},& n_\kappa&=\frac{\partial \kappa}{\partial \mu} , & n_{\bar \kappa}&=\frac{\partial\bar \kappa}{\partial \mu} , & n_{\gamma} = \frac{\partial \gamma}{\partial \mu} ,\nonumber \\ 
    \beta_0&=\frac{\partial^2 K_0}{\partial \mu^2}\, , & \beta_\kappa&=\frac{\partial^2 \kappa}{\partial \mu^2} , & \beta_{\bar \kappa}&=\frac{\partial^2 \bar \kappa}{\partial \mu^2} .  
\end{align}
Notice that in \eqref{YoungCoefChange} we included the possibility of having a pressure difference $\Delta p$ between the fluid outside and inside the membrane. It is known that the pressure difference will enter the Young-Laplace equation as shown in \eqref{YoungCoefChange}, see \cite{Armas:2020} for a derivation of this particular contribution using the Newton-Cartan sub-manifolds formalism. For completeness, we will now use the axisymmetric configuration discussed in section IV.B to write the explicit form of the differential Young-Laplace equation \eqref{YoungCoefChange} in terms of the function $\phi(\rho)=\sin \psi(\rho)$

\begin{align}\label{YoungLaplaceFunctional}
    &\kappa \left(\phi^2 -1 \right)\phi''' + \kappa \left( \phi \phi' + \frac{2(\phi^2-1)}{\rho}  \right) \phi'' - \frac{\kappa}{2}\left( \phi'^2 -  \frac{3\phi \phi'}{\rho} + \frac{3\phi^2-2}{\rho^2} \right) \phi' \nonumber \\
    &+ \frac{\kappa}{2} \left( \frac{\phi^2 -2}{\rho^3} \right)\phi  + \gamma \left( \phi' + \frac{\phi}{\rho} \right) + \frac{K_0 \kappa}{2} \left( K_0 + \frac{4 \phi}{\rho} \right) \phi' +  \frac{\kappa K_0^2}{2} \frac{\phi}{\rho} \nonumber \\
    & + \frac{n_0 \kappa \omega^2}{4} \left( \rho \phi (K_0 +2 \phi') +3 \phi^2 -2 \right) + \frac{n_{\bar \kappa}\omega^2 }{2}\left[ (3\phi^2 -1 ) \phi' + \frac{(\phi^2 -1)\phi}{\rho} \right]  \nonumber \\
    &+\frac{n_\gamma \omega^2  }{4} \rho \phi+ n_\kappa \omega^2 \left[ \frac{\rho(\phi^2 -1 ) \phi''}{2}+ \frac{\phi'}{8}\left(\rho \phi(3\phi'+4 K_0 )+ 12 \phi^2 -8 \right) \right.
    \nonumber \\
    & \left. + \frac{\phi}{8}\left(\frac{\phi^2}{\rho} + 6 K_0 \phi + \rho K_0^2  \right) - \frac{K_0}{2} \right] + \frac{\omega^4 n_0 n_\kappa  }{8} \rho^2 (\phi^2 -1) \nonumber \\
    &+ \frac{\omega^4 \beta_\kappa \rho}{16} \left( \rho \phi^2 (K_0 + \phi') - \rho (K_0 +\phi') + \phi^3- \phi  \right)+ \frac{\omega^4 \beta_{\bar \kappa}}{8} \rho \phi(\phi^2 -1 ) \nonumber \\ 
    &+ \frac{\omega^4 \beta_0 \kappa \rho^2(\phi^2-1)}{16}  = \Delta p. 
\end{align}

Equation \eqref{YoungLaplaceFunctional}, for $\omega=0$ and constant thermodynamic coefficients  $\{\gamma,\kappa, K_0, \bar \kappa \}$, has been known for quite some time \cite{Hiroyoshi1996}. Under those conditions, equation \eqref{YoungLaplaceFunctional} admits some analytic solutions such as the sphere, the torus, and the biconcave disc \cite{Tu2014}. Note that when $\bar \kappa$ is taken to be constant, it does not contribute to the Young-Laplace equation because its associated term in the equilibrium partition function becomes a constant term determined by the topology of the membrane\footnote{ Gauss-Bonnet theorem tells us that in two spatial dimensions, the curvature $\mathcal{R}$ is a total derivative, whose integral over the spatial dimensions is the Euler characteristic, which is fully determined by the topology of the surface. If the coefficient in front of the $\mathcal{R}$ term in the equilibrium partition function is a constant, then its contribution can be integrated out, and it will therefore not contribute to the Young-Laplace equation.}. In the rest of the section, we will look at some solutions to \eqref{YoungLaplaceFunctional} assuming different functional dependencies for the thermodynamic coefficients to explore some of the shapes the equation allows.
In what follows, we are going to need the equations of state for the thermodynamic coefficients. We will assume all of them are regular functions of the chemical potential $\mu$ and admit the generic expansion 
%%%%
\begin{align}\label{expGamma}
    \gamma(\mu) & = \gamma^{(0)}+ n_\gamma^{(0)} \mu + \frac{1}{2} \beta_\gamma^{(0)} \mu^2 + ...  ,\\
    \kappa(\mu) & =\kappa^{(0)} + n_\kappa^{(0)}\mu + \frac{1}{2} \beta_\kappa^{(0)} \mu^2 + ... , \\ 
    K_0(\mu) & =K_0^{(0)} + n_0^{(0)}\mu + \frac{1}{2} \beta_0^{(0)} \mu^2 + ... ,  \\
    \bar \kappa(\mu) & =\bar \kappa^{(0)} + n_{\bar \kappa}^{(0)}\mu + \frac{1}{2} \beta_{\bar \kappa}^{(0)} \mu^2 + ... , \label{expKappaBar}
\end{align}
%%%
where $\{\gamma^{(0)},n^{(0)}_\gamma,\beta_\gamma^{(0)}, \kappa^{(0)},n_\kappa^{(0)},\beta_\kappa^{(0)}, K_0^{(0)},n_0^{(0)},\beta_0^{(0)},\bar \kappa^{(0)}, $ $ n_{\bar \kappa}^{(0)},\beta_{\bar \kappa}^{(0)}  \}$ are all constants \footnote{In principle, these coefficients can be functions of the temperature. However,  in equilibrium, the temperature must necessarily be constant} and where ``...'' denotes higher order terms in the chemical potential that will eventually be taken to be zero and will not play any role in our discussion.
%%%

\vspace{2mm}

We will be interested in looking at closed membrane solutions. For solutions with the topology of the sphere, we will have the boundary conditions $\phi(0)=0$ and $\phi(r)=-1$, where $r$ is the maximum width of the membrane as measured in the ambient space and where the condition at $\rho=0$ ensures the solution has no kinks. For toroidal solutions, the boundary conditions will be $\phi(r_+-r_-)=1$ and $\phi(r_++r_-)=-1$, where $r_-$ is the torus minor radius and $r_+$ is the torus major radius.

For all solutions that we looked at in this work, we have assumed thermodynamic relations linear in $\mu$, namely only $\{\gamma^{(0)}, n^{(0)}_\gamma, \kappa^{(0)},n^{(0)}_\kappa, K^{(0)}_0, n^{(0)}_0,$
$\bar \kappa^{(0)},n^{(0)}_{\bar \kappa} \}$ have the possibility of being non-vanishing.

\subsection{Spherical Solution}

We will denote a spherical solution as $\phi_S(\rho)$. Such solution is given by
%%%
\begin{align}\label{sphereSol}
    \phi_S(\rho) = - \frac{\rho}{r}, 
\end{align}
%%%
with $r$ the radius of the sphere. The function \eqref{sphereSol} is a solution to \eqref{YoungLaplaceFunctional} when 
%%%
\begin{align}
  \gamma(\mu) & = \gamma^{(0)} + \left[- \frac{K^{(0)}_0 n^{(0)}_\kappa }{2} + \frac{3 K^{(0)}_0 n^{(0)}_\kappa }{r} \right. \nonumber  \\ &  \hphantom{\gamma^{(0)} ++++} \left. -\frac{4(n_{\kappa}^{(0)}+n_{\bar \kappa}^{(0)})}{r^2} \right] \mu , 
 \label{SphereSurfTension}
 \\
  \kappa(\mu) & = \kappa^{(0)} + n^{(0)}_\kappa \mu  ,\\
  K_0(\mu) & = K^{(0)}_0 , \\
  \bar \kappa(\mu) &= \bar \kappa^{(0)} + n^{(0)}_{\bar \kappa} \mu ,
\end{align}
%%%
and when the following constraint equation is satisfied
%%%%
\begin{align}
    r^2 \Delta \tilde p   + r \left( \left(K^{(0)}_0\right)^2 \kappa^{(0)} + 2 \tilde \gamma^{(0)}   \right) - 2 K^{(0)}_0 \kappa^{(0)} =0 ,
    \label{SpheraConstraint}
\end{align}
%%%
where we defined the shifted pressure $\Delta \tilde p$ and the shifted surface tension $\tilde \gamma^{(0)}$ as
%%%
\begin{align}
   \Delta \tilde p & =  \Delta p + \frac{1}{2}K^{(0)}_0 n_\kappa^{(0)}\omega^2 , \\
   \tilde \gamma^{(0)} &  = \gamma^{(0)} - \frac{1}{2} \left(n^{(0)}_\kappa + n_{\bar \kappa}^{(0)} \right)  \omega^2 . 
\end{align}
\textcolor{red}{}
%%%%

Notice that the sphere is a solution for $\omega = 0$ \cite{Armas:2020}. 
For a rotating membrane, the Young-Laplace equation (\ref{YoungLaplaceFunctional}) admits a spherical solution under specific conditions.
Given a linear dependence on $\mu$ of the elastic moduli, and a constant intrinsic curvature $K_0^{(0)}$, the surface tension relation (\ref{SphereSurfTension}) will determine the sphere radius, as long as the pressure difference is set according to (\ref{SpheraConstraint}).

\subsection{Biconcave Disc Solution}

We will denote a biconcave disc solution as $\phi_D(\rho)$. Such solution is given by
%%%%
\begin{align}\label{solDisc}
    \phi_D(\rho) = - \rho \left( \frac{1 - r K^{(0)}_0 \ln r}{r} + K^{(0)}_0 \ln \rho \right)  , 
\end{align}
%%%
where $r$ is the maximum horizontal length attained by the discoid. The function \eqref{solDisc} is a solution to \eqref{YoungLaplaceFunctional} when
%%%%
\begin{align}
  \gamma(\mu)&=0 , \\
   \kappa(\mu)&=\kappa^{(0)} + n_{\kappa}^{(0)} \mu , \\
    K_0(\mu) & = K^{(0)}_0 , \\ 
    \bar \kappa(\mu) &= \bar \kappa^{(0)} - n_{\kappa}^{(0)} \mu , 
\end{align}
and where the pressure difference takes the value
%%%%
\begin{align}
    \Delta p = \frac{n_{\kappa}^{(0)}  \omega^2}{2}. 
\end{align}

%%%

As in the case of $\omega=0$, the discoid is a solution of zero surface tension \cite{Armas:2020}, but in contrast, it needs a nonzero pressure difference proportional to $\omega^2$. 
What seems very restrictive in the discoid solution is the relations for the elastic moduli, where their linear rate of change with $\mu$ differs only in their sign.

\subsection{Torus Solution}

We will denote a toroidal solution as $\phi_T(\rho)$
%%%%
\begin{align}\label{solTorus}
    \phi_T(\rho)=- \frac{\rho}{r_-} + \frac{r_+}{r_-} \, , 
\end{align}

where $r_-$ is the minor radius of the torus and $r_+$ is the mayor radius of the torus.  The function \eqref{solTorus} is a solution to \eqref{YoungLaplaceFunctional} when 
%%%%
\begin{align}
    \gamma(\mu) &= \left[ \frac{K^{(0)}_0 \kappa^{(0)} \left(4- K^{(0)}_0 r_-  \kappa^{(0)} \right)  }{2r_-} \right] \\ \nonumber &- \left[ \frac{n^{(0)}_{\bar \kappa} \left(2 -3 K^{(0)}_0 r_-   \right) \omega^2}{6 \left( K^{(0)}_0 r_- - 2 \right) }  \right] \\ \nonumber &- \frac{4 n_{\bar \kappa}^{(0)} \left( K^{(0)}_0 r_- - 1 \right)  }{3 r_-^2} \mu  \, , \\
    \kappa(\mu) &= \kappa^{(0)} + \frac{8 n_{\bar \kappa}^{(0)}}{3 \left( K^{(0)}_0 r_- - 2 \right)} \mu \, , \\
    K_0(\mu) &= K^{(0)}_0 \, , \\
    \bar \kappa (\mu) &= \bar \kappa^{(0)} + n^{(0)}_{\bar \kappa} \mu 
\end{align}
where $r_-$ and $r_+$ are not independent and are related through
%%%
\begin{align}
    r_+ = \sqrt{2} r_- \, ,
    \label{RadiiTorus}
\end{align}
%%%%
and where the pressure is constrained to take the value 
%%%%
\begin{align}
    \Delta p = \frac{2}{3 r_-^3} \left[ \frac{n^{(0)}_{\bar \kappa} r_- \left(3 - 2 K^{(0)}_0 r_- \right)\omega^2}{K^{(0)}_0 r_- - 2}- 3 K^{(0)}_0 \kappa^{(0)} \right] \, .
\end{align}

The torus is the third solution that is also a solution at $\omega =0$, with its radii that relate through (\ref{RadiiTorus}).
Nonetheless, the dependence of both $\gamma$ and $\kappa$ on $r_-$, imposes severe restrictions on the membrane material properties.

\subsection{ Biconcave disc compatible membranes}

\begin{center}
\begin{figure}[h!]
    \centering
      \includegraphics[scale=0.28]{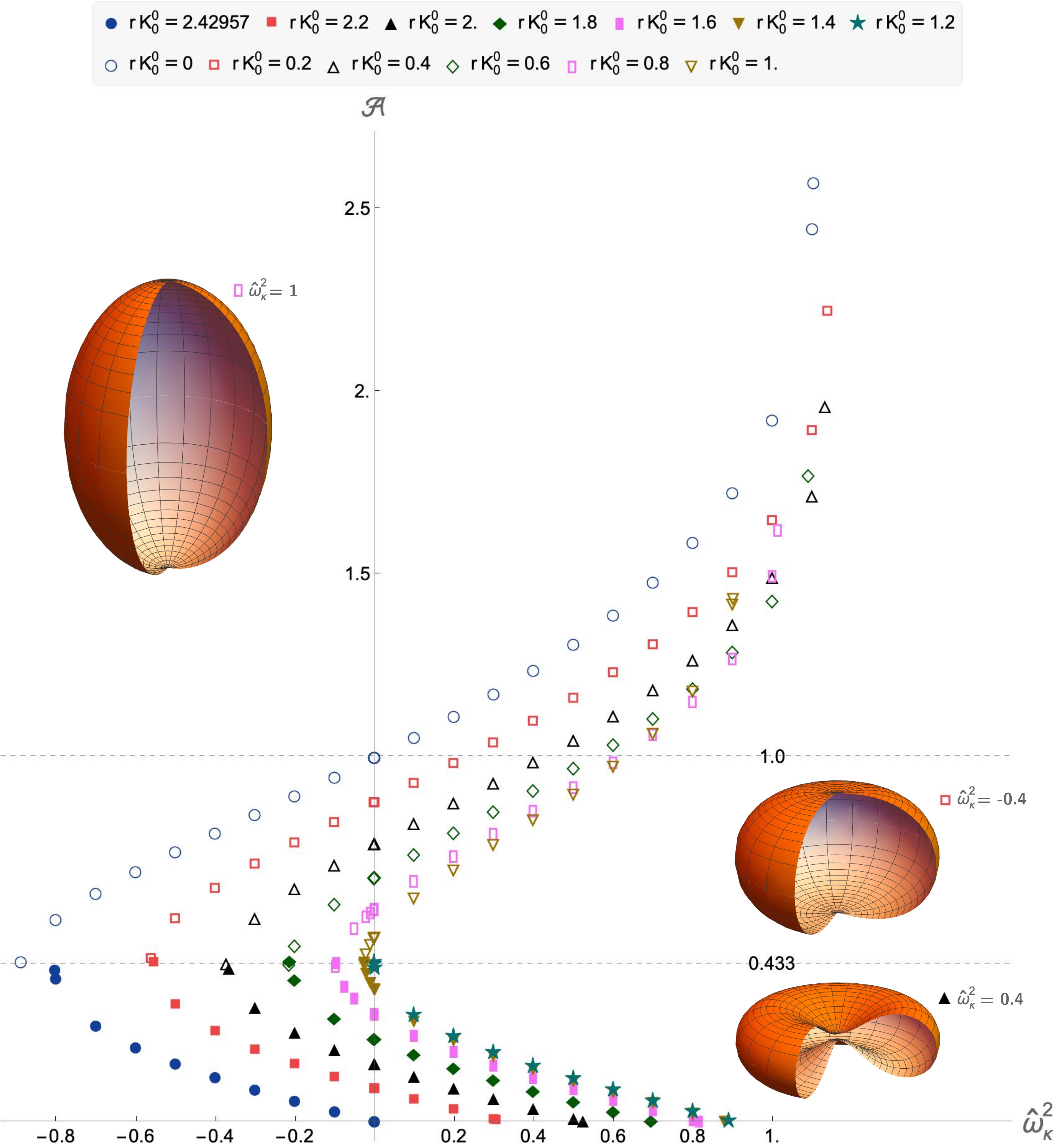}
    \caption{
      Numerical solutions to the Young-Laplace equations (\ref{YoungLaplaceFunctional}) are presented in terms of the membrane's aspect ratio $\mathcal{A}$.
     Notice that by its definition (\ref{omegagorro}), $\hat{\omega}_\kappa^2$ can take negative values.
     Characteristic shapes are included in the figure. As $\mathcal{A}$ goes from zero to approximately $0.433$, the biconcave-like shapes of the membranes start with $z_0=0$ (suggesting limiting parameter values for pinch off and topological transitions), and near $\mathcal{A}=0.433$ the membranes start looking more like oblate spheroids. The oblate spheroid resemblance is maintained above $\mathcal{A}\approx 0.433$ and up to $\mathcal{A}=1.0$.  Above $\mathcal{A}=1.0$ prolate spheroid-like shapes are adopted.
    }
    \label{fig:PlotKappa}
\end{figure}
\end{center}

We will now look at some particular numerical solutions to \eqref{YoungLaplaceFunctional}. Solutions with the topology of the sphere and with thermodynamic coefficients with at most a linear dependence on the chemical potential, namely 
\begin{align}
    \gamma(\mu) &= \gamma^{(0)} + n^{(0)}_\gamma \mu\, , \\
    \kappa(\mu) &= \kappa^{(0)} + n^{(0)}_\kappa \mu \, , \\ 
    K_0(\mu) &= K^{(0)}_0 + n^{(0)}_0 \mu \, , \\
    \bar \kappa(\mu) &= \bar \kappa^{(0)} + n^{(0)}_{\bar \kappa} \mu \, , 
\end{align}
%%%
We want to find a family of solutions that are continuously connected to the $\omega=0$ biconcave disc solution. For this to be the case near $\omega=0$, the solution must have an asymptotic behavior near $\rho=0$ that is compatible with the biconcave disc solution, namely, we will ask for the numerical solution to be compatible with the asymptotic behavior
%%%
\begin{align}\label{discCompatibility}
    \phi(\rho) = - K^{(0)}_0 \rho \left(c_1 + \ln \rho \right) + \mathcal{O}\left(\rho^2 \right) \, , 
\end{align}
%%%
for some constant $c_1$. For $\phi(\rho)$ to be compatible with \eqref{discCompatibility} the following constraints should hold
%%%%
\begin{align}\label{constraint1}
    \gamma^{(0)} &= \frac{\omega^2}{2} \left(n^{(0)}_\kappa + n^{(0)}_{\bar \kappa} \right) \, , \\
    \Delta p &= \frac{\omega^2}{2} \left( K^{(0)}_0 n^{(0)}_\kappa - \kappa^{(0)} n^{(0)}_0 \right) \, . \label{constraint2}
\end{align}
%%%
Constraints \eqref{constraint1} and \eqref{constraint2} reduce the number of independent thermodynamic coefficients to $\{K^{(0)}_0, \kappa^{(0)}, n^{(0)}_\kappa, n^{(0)}_{\bar \kappa}, n^{(0)}_0 \}$ with $\{K^{(0)}_0, \kappa^{(0)} \}$ associated to the $\omega=0$ solution and $\{n^{(0)}_\kappa, n^{(0)}_{\bar \kappa}, n^{(0)}_0\}$ associated to the solution once vorticity is turned on. The exploration of the solutions in this reduced parameter space becomes simpler, and it is the reason we choose to explore this class of solutions. Even then, the full exploration of the parameter space becomes challenging, and we will restrict ourselves to exploring a subspace of it. In particular, we will focus on the following two cases:
%%%
\begin{align}
    \text{case 1:}& \qquad n^{(0)}_0=0 \, , \qquad n^{(0)}_{\bar \kappa}=0  \, , \\
    \text{case 2:}& \qquad n^{(0)}_0=0 \, , \qquad n^{(0)}_{\kappa} =0\,, 
\end{align}
%%%
In the first case, all thermodynamic parameters are assumed constant with the exception of the bending modulus. In this case, a nonzero surface tension and a nonzero pressure difference are assumed.  In the second case, only the effect of a non-constant Gaussian modulus is studied. In this case, a nonzero surface tension and zero pressure difference are assumed. %%%%
Both cases have similar qualitative behavior while being physically distinct, as one of them requires the presence of a pressure difference. To aid in the discussion, we will define the dimensionless squared vorticities $\hat \omega^2_{\kappa/ \bar \kappa}$ as
%%%
\begin{align}
    \hat \omega^2_{\kappa} &\equiv     \frac{r^2 n^{(0)}_\kappa}{\kappa^{(0)}}  \omega^2  \, , 
    \label{omegagorro}
    \\
    \hat \omega^2_{\bar \kappa} &\equiv  \frac{r^2 n^{(0)}_{\bar \kappa}}{\kappa^{(0)}} \omega^2 \, , 
\end{align}
%%%%
These parameters correspond to the dimensionless vorticities square, where the maximum radius $r$, $\kappa^{(0)}$, and $n^{(0)}_{\kappa/\bar \kappa}$ were used for this purpose. Note that despite the notation, $\hat{\omega}_{\kappa/\bar \kappa}^2$ can take negative values when $n^{(0)}_{\kappa/\bar \kappa}/\kappa^{(0)}$ takes negative values. 
For both cases ($n^{(0)}_\kappa=0$ or $n^{(0)}_{\bar \kappa}=0$), we find two distinct families of solutions that we classify using the dimensionless parameter $rK^{(0)}_0$. It is also convenient to use the height $z_0 = - \int_0^R d \rho \frac{\phi(\rho) }{\sqrt{1 - \phi(\rho)^2 }}$ to define the aspect ratio $\mathcal{A}$ as 
%%%%
\begin{align}
    \mathcal{A}= \frac{z_0}{r} \,  . 
\end{align}
%%%%
Using this definition for $\mathcal{A}$, we will call $\mathcal{A}_0$ the aspect ratio at $\hat{\omega}_{\kappa/\bar \kappa}^2=0$ for any particular given $r K^{(0)}_0$. We will use $\mathcal{A}$ to characterize the qualitative features of the solutions. Using $r K^{(0)}_0$ the following two distinct families are found:
%%%%%

\begin{enumerate}[label=(\alph*)]
    \item $0\leq r K^{(0)}_0 < 1.2$ 
%%%

    The lower limit $r K^{(0)}_0=0$ at $\hat{\omega}_{\kappa/\bar \kappa}^2=0$ 
    corresponds to the limit in which the biconcave disc solution becomes a sphere; consequently $\mathcal{A}_0=1$ for this case. 
    The upper boundary $r K^{(0)}_0=1.2$ corresponds to a biconcave disc solution with an aspect ratio of $\mathcal{A}_0\approx 0.43$. 
    In figures \ref{fig:PlotKappa} and \ref{fig:PlotBarKappa}
    a plot for $\mathcal{A}$ vs $\hat \omega_{\kappa}^2$ and $\mathcal{A}$ vs $\hat \omega_{\bar \kappa}^2$ are shown respectively. These plots are very similar; therefore, figure \ref{fig:PlotBarKappa} is presented in Appendix \ref{Appendix}.  

    %%%%

    For all $r K^{(0)}_0$ that we looked at and for $\hat \omega_{\kappa/\bar \kappa}^2<0$,  the aspect ratio decreases compared to the $\hat \omega_{\kappa/\bar \kappa}^2=0$ solution, namely $\mathcal{A}<\mathcal{A}_0$. The shape of the solutions goes from a sphere to a shape that resembles a biconcave disc with aspect ratio $\mathcal{A}$. In between these shapes, and for most of this regime, the membrane takes a shape that resembles an oblate spheroid, see figure \eqref{fig:PlotKappa} for a characteristic example of the shapes found in this regime. It is important to note that while the solution has the visual features of an oblate spheroid, it does not follow the exact analytic form of one. 
    
    These solution follow a decreasing trend in $\mathcal{A}$ as $\hat \omega_{\kappa/\bar \kappa}^2$ decreases until some lower critical value $(\hat \omega^2_{\kappa/\bar \kappa})_{\text{b-crit}}$ that depends on  $r K^{(0)}_0$, after which we were unable to find solutions that are continuously connected to the trend line connecting the solutions for $(\hat \omega^2_{\kappa/\bar \kappa})_{\text{b-crit}} < \hat \omega_{\kappa / \bar{\kappa}}^2\leq 0$. Note that $(\hat \omega^2_{\kappa/\bar \kappa})_{\text{b-crit}}$ and its corresponding critical aspect ratio $\mathcal{A}_{\text{b-crit}}$ are different for each $r K^{(0)}_0$, although we can note that $\mathcal{A}_{\text{b-crit}}$ is at most within $3\%$ of the aspect ratio for $r K^{(0)}_0=1.2$ at $\hat \omega_{\kappa/\bar \kappa}^2=0$. 
    %%%%

    For all $r K^{(0)}_0$ that we looked at and for $\hat \omega_{\kappa/{\bar \kappa}}^2>0$, the aspect ratio increases compared to the solution $\hat \omega_{\kappa/ \bar \kappa}^2=0$, namely $\mathcal{A}> \mathcal{A}_0$. The shape of the solutions becomes similar to that of a prolate spheroid, see figure \ref{fig:PlotKappa} for a characteristic example of the shapes found in this regime. It is important to note that while the solution resembles a prolate spheroid, it does not follow the exact analytic form of one. These solutions follow an increasing trend in $\mathcal{A}$ as $\hat \omega_{\kappa / \bar \kappa}^2$ increases until a critical $(\hat \omega^2_{\kappa/\bar \kappa})_{\text{u-crit}}$ after which we were unable to find solutions that are continuously connected to the trend line connecting the solutions for $0\leq \hat \omega_{\kappa /\bar \kappa}^2 < (\hat \omega^2_{\kappa/\bar \kappa})_{\text{u-crit}} $.
    
    \item $1.2 \leq r K^{(0)}_0 \lessapprox 2.42957 $
    
    The lower limit $r K^{(0)}_0 = 1.2$ at $\hat \omega_{\kappa/ \bar \kappa}^2=0$ corresponds to a biconcave disc solution with an aspect ratio of $\mathcal{A}_0 \approx 0.43$. The upper limit $ r K^{(0)}_0 \approx 2.42957 $ corresponds to a biconcave disc solution with approximately zero height, namely $\mathcal{A}_0 \approx 0$. In figure \ref{fig:PlotKappa} and \ref{fig:PlotBarKappa} a plot for $\mathcal{A}$ vs $\hat \omega_{\kappa}^2$ and $\mathcal{A}$ vs $\hat \omega_{\bar \kappa}^2$ is shown respectively. Note that the plot contains both families, the one currently being discussed and the previously discussed one. 
    %%%%%

    For all $r K^{(0)}_0$ that we looked at and for $\hat \omega_{\kappa / \bar \kappa}^2<0$, the aspect ratio increases compared to the $\hat \omega_{\kappa/\hat \kappa}^2=0$ solution, namely $\mathcal{A}>\mathcal{A}_0$. These solutions follow an increasing trend in $\mathcal{A}$ as $\hat \omega_{\kappa/\bar \kappa}^2 $ decreases until a critical $(\hat \omega^2_{\kappa/\bar \kappa})_{\text{b-crit}}$ after which we were unable to find solutions that are continuously connected to the trend line connecting the solutions for $(\hat \omega^2_{\kappa/\bar \kappa})_{\text{b-crit}} \leq \hat \omega_{\kappa / \bar \kappa}^2 \leq 0$. Note that $(\hat \omega^2_{\kappa/\bar \kappa})_{\text{b-crit}}$ and its corresponding critical aspect ratio $\mathcal{A}_{b-crit}$ are different for each $r K^{(0)}_0$, although we can note that $\mathcal{A}_{b-crit}$ is at most within 4\% of the aspect ratio for $r K^{(0)}_0 = 1.2$ at $\hat \omega_{\kappa / \bar \kappa}^2=0$.   
    The shape of the solutions remains visually that of a biconcave disc, see figure \ref{fig:PlotKappa} for a characteristic example of the shapes found in this regime. Note that although the shape resembles a biconcave disc, it is not an exact analytical solution such as \eqref{solDisc}. 
    %%%%

    For all $r K^{(0)}_0$ that we looked at and for $\hat \omega_{\kappa / \bar \kappa}^2>0$, the aspect ratio decreases compared to the $\hat \omega_{\kappa / \bar \kappa}^2=0$ solution, namely $\mathcal{A}<\mathcal{A}_0$. These solutions follow a decreasing trend in $\mathcal{A}$ as $\hat \omega_{\kappa / \bar \kappa}^2$ increases until a critical $(\hat \omega^2_{\kappa/\bar \kappa})_{\text{u-crit}}$ at which the aspect ratio becomes zero. This suggests a critical rotational speed at which the membrane might suffer a topological change and become a torus, similar to what is observed in active membranes due to an increasing activity \cite{hoffmann}. Note that the shapes of the membranes in this regime are still visually that of a biconcave disc. 
\end{enumerate}

 Notice that the choice of axis in plots \ref{fig:PlotKappa} and \ref{fig:PlotBarKappa}, allowed us to present the different types of approximate solutions to membrane shapes in terms of the aspect ratio $\mathcal{A}$.
Points in these plots, even the ones with the same symbol (the same value of $rK_0^{(0)}$) are not necessarily connected, in the sense that two solutions on the same $r K_0^{(0)}$ line do not represent different shapes of the same membrane subject to a different rotation given by $\omega$.
 For example, if we were interested in the response to rotation of a closed membrane with some given material properties and that is constrained to keep its volume constant, we would need to compute and present the solutions in a different manner. 

Prolate membrane solutions, $\mathcal{A}>1$, were found for $\hat \omega^2_{\kappa/\bar \kappa}>0$, although it seems counterintuitive that a spinning membrane can acquire a prolate shape.
Notice that, considering $\kappa^{(0)} >0$, the obtained prolate solutions have either $n_{\kappa}^{(0)}>0$ and $n_{\bar{\kappa}}^{(0)}=0$, or $n_{\kappa}^{(0)}=0$ and $n_{\bar{\kappa}}^{(0)}>0$.
This means that their elastic moduli are growing functions of $\mu$, and therefore the rigidity of such membranes increases radially when rotating, given the radial dependence of $\mu$ (\ref{chemicalPotential}).

Particularly interesting is the case of $K_0^{(0)}=0$, the open blue circles in figures \ref{fig:PlotKappa} and \ref{fig:PlotBarKappa}.
These solutions, for $\mathcal{A}>0$ can represent the shapes a membrane with given elastic moduli (that grows with $\mu$), and that at rest takes a spherical shape, can acquire as it rotates faster.
On the other hand, the branch $\mathcal{A}<0$ corresponds to radially decreasing elastic moduli that become less rigid with the radius. This branch can be interpreted as the membrane, which is spherical at rest, tendency to become oblate as it rotates, eventually taking a shape that resembles a biconcave disc.

\section{Conclusions}\label{Sec:Con}

We have successfully used the formalism presented in \cite{Armas:2020} to construct the equilibrium partition function and the constitutive relations for co-dimension one NC membranes embedded in flat NC spacetime with both finite temperature and chemical potential up to second order in a derivative expansion.
The equilibrium constraints imply that the temperature must be constant and the chemical potential must be related to the acceleration. 
%%%%
The shape of the membrane in its equilibrium state will be determined by the generalized Young-Laplace equation \eqref{YoungLaplace}. 
We wrote down this equation explicitly for a particular set of constitutive relations in the presence of constant vorticity. 
We would like to point out that in \cite{Anand2015} a relation between linear chemo-elasticity and linear poroelasticity was established, which means the presence of a mass chemical potential can be interpreted as having an underlying effective porous medium. This will allow us to eventually extend the range of applications of the resulting work in the context of finite temperature poroelastic membranes at hydrostatic equilibrium.

As happens in the case of a static closed membrane \cite{Armas:2013}, spheres, tori, and biconcave discoids are solutions to the Young-Laplace equations when $\omega \neq 0$, as long as the elastic moduli and spontaneous curvature are linear function of the chemical potential $\mu$ and specific relations are met by the surface tension and pressure with the elastic moduli and spontaneous curvature parameters.
Additionally, it is possible to find numerically a family of axisymmetric solutions with material conditions that allow us to continuously connect them to the $\omega=0$  biconcave disc analytic solution.

\section*{Acknowledgement}

This work was supported by Universidad Nacional Aut\'onoma de M\'exico through UNAM-DGAPA's posdoctoral fellowship program "Programa de Becas Posdoctorales en la UNAM."

\appendix

\section{Appendix}

In here we show the variation of each of the terms appearing in the equilibrium partition function
\subsection{Variation of $\chi$}

The variation coming from the $\chi(T,\mu)$ term is explicitly given by
\begin{align}
    \delta \mathcal{F}_{\chi} &= \int d^{d-1} \sigma \hat e \left[  \frac{1}{2}\left( \chi h^{ab}  + n u^a u^b \right) \delta \bar h_{ab} - \left( \chi \hat \upsilon^c + \left( \epsilon + \chi + \frac{n}{2}\bar u^2  \right)u^c  \right) \delta \tau_c  \right].
\end{align}
%%%
The contribution of this variation to the currents is given by
%%%
\begin{align}
    \left(\mathcal{T}_S\right)^c_{\chi} & =- \chi \hat \upsilon^c - \left( \epsilon + \chi + \frac{n}{s} \bar u^2  \right)u^c ,\\
    \left( \mathcal{T}_S \right)^{ab}_\chi  & = \chi h^{ab} + n u^a u^b , \\
    D^{cd}_\chi & =0.
    \end{align}

\subsection{Variations of order one scalars}

The variation of $F_{(1)}=K_{ab}h^{ab}$ is 
%%%
\begin{align}
    \delta F_{(1)} &= - \left( K_{a b} h^{ac} h^{b d} \right) \delta h_{ c d} + \left( 2 \hat \upsilon^a h^{c b} K_{ab} \right)\delta \tau_c + \left( h^{cd} \right)\delta K_{cd} ,
\end{align}
%%%
contributing to a variation of the equilibrium partition function given by
%%%%
\begin{align}
    \delta \mathcal{F}_{ F_{(1)} } &= \int d^{d-1} \sigma \hat e \left[ \frac{1}{2} \left( \xi^{(1)cd}_1 F_{(1)} - 2 a_1^{(1)} h^{ac} h^{bd} K_{ab} \right) \delta \bar h_{cd} \right] \\
    &+ \int d^{d-1} \sigma \hat e \left[ \left(2 a_1^{(1)} \hat \upsilon^a h^{bc} K_{ab} -\zeta^{(1)c}_1 F_{(1)} \right)\delta \tau_c + \left( a_1^{(1)} h^{cd} \right) \delta K_{cd} \right] .\nonumber 
\end{align}
%%%%
The contribution of this variation to the currents is given by
%%%
\begin{align}
     \left(\mathcal{T}_S\right)^c_{F_{(1)}} & =-a_1^{(1)} K \hat \upsilon^c - \left[s_1^{(1)}T + n_1^{(1)} \left(\mu + \frac{1}{2} \bar u^2 \right) \right] K u^c + 2 a^{(1)}_1 \hat \upsilon^a h^{bc} K_{ab} , \\
    \left( \mathcal{T}_S \right)^{cd}_{F_{(1)}} & = a^{(1)}_1 K h^{cd} + n^{(1)}_1 K u^c u^d - 2 a^{(1)}_1 h^{ac}h^{bd} K_{ab} , \\
    D^{cd}_{F_{(1)}} & =a_1^{(1)} h^{cd}.
\end{align}
%%%%
The variation of $F_{(2)}=K_{ab}u^a u^b$ is
%%%%
\begin{align}
    \delta F_{(2)} &= - \left( 2 K_{ab} u^a u^b u^c\right) \delta \tau_c + \left(u^c u^d \right) \delta K_{cd} ,
\end{align}
%%%%%
contributing to a variation of the equilibrium partition function given by
%%%
\begin{align}
\delta \mathcal{F}_{F_{(2)}} &= \int d^{d-1} \sigma \hat e \left[ \frac{1}{2}\left(\xi^{(1)cd}_2 F_{(2)} \right)\delta \bar h_{cd} - \left( \zeta^{(1)c}_2 F_{(2)} + 2 a_2^{(1)} K_{ab}u^a u^b u^c \right) \delta \tau_c   \right] \\
& + \int d^{d-1} \sigma \hat e \left(a^{(1)}_{2} u^c u^d\right) \delta K_{cd}. \nonumber 
\end{align}
%%%
The contribution of this variation to the currents is given by
%%%%
\begin{align}
    \left(\mathcal{T}_S\right)^c_{F_{(2)}} & = - a_2^{(1)} \left( a \cdot n \right)\hat \upsilon^c - \left[ s_2^{(1)} T + n_2^{(1)} \left(\mu + \frac{1}{2} \bar u^2 \right)+ 2 a^{(1)}_2 \right] \left(a \cdot n \right) u^c , \\
    \left( \mathcal{T}_S \right)^{cd}_{F_{(2)}} & = 0 , \\
    \mathcal{D}^{cd}_{F_{(2)}}& = a^{(1)}_2 u^c u^d \, .
\end{align}
\subsection{Variations of order two scalars}

The variation of $S_{(1)} = \left( K_{ab} h^{ab} \right) \left( K_{cd} h^{cd}\right)$ is 
%%%
\begin{align}
    \delta S_{(1)} = 2 F_{(1)} \delta F_{(1)}.
\end{align}
%%%
Hence, it will contribute to the variation of the equilibrium partition function as
%%%
\begin{align}
    \delta \mathcal{F}_{ S_{(1)} } &= \int d^{d-1} \sigma \hat e \left[ \frac{1}{2} \left( \xi^{(2)cd}_1 S_{(1)} - 4 a_1^{(2)} F_{(1)} h^{ac} h^{bd} K_{ab} \right) \delta \bar h_{cd} \right] \\
    &+ \int d^{d-1} \sigma \hat e \left[ \left(4 a_1^{(2)} F_{(1)} \hat \upsilon^a h^{bc} K_{ab} -\zeta^{(2)c}_1 S_{(1)} \right)\delta \tau_c \right. \nonumber \\
    & \left. + \left( 2a_1^{(2)} F_{(1)} h^{cd} \right) \delta K_{cd} \right]. \nonumber 
\end{align}
%%%%
The contribution of this variation to the currents is given by
%%%
\begin{align}
    \left( \mathcal{T}_S \right)^c_{S_{(1)}} & =- a^{(2)}_1 K^2 \hat \upsilon^c - \left[s^{(2)}_1 T + n^{(2)}_1\left( \mu + \frac{1}{2}\bar u^2 \right) \right] K^2 u^c + 4 a^{(2)}_1 K \hat \upsilon^a h^{bc} K_{ab} , \\ 
    \left( \mathcal{T}_S \right)^{cd}_{S_{(1)}} & = a^{(1)}_2 K^2 h^{cd} + n^{(2)}_1 K^2 u^c u^d -4 a^{(2)}_1 K h^{ac}h^{bd} K_{ab} , \\
    \mathcal{D}^{cd}_{S_{(1)}} &= 2 a^{(2)}_1 K h^{cd}. 
\end{align}
%%%%
The variation of $S_{(2)}=K_{ab} K_{cd} h^{ac} h^{bd}$ is
%%%
\begin{align}
    \delta S_{(2)} &= \left(-2 h^{ac} h^{bd} h^{rs} K_{ar} K_{bs}  \right) \delta \bar h_{cd} + \left( 4 \hat \upsilon^{a} h^{bc} h^{dr} K_{ad} K_{br} \right) \delta \tau_c \nonumber \\
    & + \left(2h^{ac} h^{bd} K_{ab} \right) \delta K_{cd},
\end{align}
%%%
and it will contribute to the variation of the equilibrium partition function as
%%%
\begin{align}
    \delta \mathcal{F}_{S_{(2)}} &= \int d^{d-1} \sigma \hat e \left[ \frac{1}{2} \left(\xi_2^{(2)cd} S_{(2)} - 4 a_2^{(2)} h^{ac} h^{bd} h^{rs} K_{ar} K_{bs} \right) \delta \bar h_{cd} \right] \\
    \nonumber 
    &+ \int d^{d-1} \sigma \hat e \left[ \left(4 a_2^{(2)} \hat \upsilon^a h^{bc} h^{dr} K_{ad} K_{br} - \zeta_2^{(2)c} S_{(2)} \right)\delta \tau_c \right. \nonumber \\
    & \left. + \left(2 a_2^{(2)} h^{ac} h^{bd} K_{ab} \right) \delta K_{cd} \right].
\end{align}
%%%
The contribution of this variation to the currents is given by
%%%%
\begin{align}
    \left( \mathcal{T}_S \right)^c_{S_{(2)}} & =-a^{(2)}_2 \left(K \cdot K \right) \hat \upsilon^c - \left[s^{(2)}_2 T + n^{(2)}_2 \left(\mu + \frac{1}{2} \bar u^2 \right) \right] \left(K \cdot K \right) u^c \\
    &+ 4 a^{(2)}_2 \hat \upsilon^a K_{ad} h^{dr} K_{rb} h^{bc}, \nonumber \\
    \left( \mathcal{T}_S \right)^{cd}_{S_{(2)}} &={ a^{(2)}_2 \left( K \cdot K \right) h^{cd} + n^{(2)}_2 \left(K \cdot K \right) u^c u^d -4 a^{(2)}_2 K_{ra}h^{rs} K_{sb} h^{ac} h^{bd} ,} \\
    \mathcal{D}^{cd}_{S_{(2)}}& =2 a^{(2)}_2 h^{ac} h^{bd} K_{ab} .
\end{align}
%%%
The variation of $S_{(3)}=K_{ab}K_{cd}u^a h^{bc} u^d$ is
%%%
\begin{align}
    \delta S_{(3)} &= \left(-h^{cr} h^{ds} K_{ar} K_{bs} u^a u^b \right)\delta h_{cd} \\
    &+ \left(2 \hat \upsilon^a h^{cr} K_{ab} K_{d r} u^b u^d  - 2 u^c h^{dr} K_{ad} K_{br} u^a u^b \right) \delta \tau_c \nonumber \\ 
    & + \left(2 u^{(c}h^{d)b} u^a K_{ab}  \right) \delta K_{cd}, \nonumber
\end{align}
%%%
and it will contribute to the variation of the equilibrium partition function as
%%%%
\begin{align}
    \delta \mathcal{F}_{S_{(3)}} &= \int d^{d-1} \sigma \hat e \left[ \frac{1}{2} \left(\xi^{(2)cd}_3 S_{(3)} - 2 a_3^{(2)} h^{cr} h^{ds} K_{ar} K_{bs} u^a u^b \right)\delta \bar h_{cd} \right] \\
    &+ \int d^{d-1} \sigma \hat e \left[ \left(2 a_3^{(2)} \hat \upsilon^a h^{cr} K_{ab} K_{dr} u^b u^d- 2 a_3^{(2)}u^c h^{dr} K_{ad} K_{br} u^a u^b \right. \right. \nonumber \\ 
    & \left. \left. -\zeta^{(2)c}_{3}S_{(3)}\right)\delta \tau_c \right] + \int d^{d-1} \sigma \hat e \left[  \left(2 a_3^{(2)} u^{(c} h^{d)b}u^a K_{ab} \right) \delta K_{cd} \right]. \nonumber
\end{align}
%%%
The contribution of this variation to the currents is given by
%%%%
\begin{align}
   \left(\mathcal{T}_S\right)^c_{S_{(3)}} & = - a^{(2)}_3 \left( n \cdot \Omega \right)^2 \hat \upsilon^c - \left[ s_3^{(2)} T + n^{(2)}_3 \left(\mu + \frac{1}{2} \bar u^2 \right) \right] \left( n \cdot \Omega \right)^2 u^c  \\ \nonumber
  & +2 a^{(2)}_3 \hat \upsilon^a h^{cr} K_{ab} K_{dr} u^b u^d - 2 a^{(2)}_3 u^c h^{dr} K_{ad} K_{br} u^a u^b, \\ 
  \left(\mathcal{T}_S\right)^{cd}_{S_{(3)}} & = a^{(2)}_3 \left(n \cdot \Omega \right)^2 h^{cd} + n_3^{(2)} \left(n \cdot \Omega \right)^2 u^c u^d - 2 a_3^{(2)} h^{cr}h^{ds} K_{ar} K_{bs} u^a u^b , \\
  \mathcal{D}^{cd}_{S_{(3)}} & =2 a^{(2)}_3 u^{(c}h^{d)b} u^a K_{ab} . 
\end{align}
%%%%%
The variation $S_{(4)}=\left(K_{ab}u^a u^b\right) \left(K_{cd}u^c u^d \right)$ is
%%%
\begin{align}
    \delta S_{(4)} &= 2 F_{(2)} \delta F_{(2)} ,
\end{align}
%%%
meaning it will contribute to the variation of the equilibrium partition function as 
%%%%
\begin{align}
    \delta \mathcal{F}_{S_{(4)}} &= \int d^{d-1} \sigma \hat e \left[ \frac{1}{2} \left( \xi^{(2)cd}_4  S_{(4)}  \right)\delta \bar h_{cd} \right. \\ 
    & \left. -\left(\zeta_4^{(2)c} S_{(4)} + 4 a_4^{(2)} F_{(2)} K_{ab}u^a u^b u^c \right)\delta \tau_c \right] \nonumber \\
    &+ \nonumber \int d^{d-1} \sigma \hat e \left(2 F_{(2)} a_4^{(2)} u^c u^d \right) \delta K_{cd}.
\end{align}
%%%%
The contribution of this variation to the currents is given by
%%%%
\begin{align}    \left(\mathcal{T}_S\right)^c_{S_{(4)}}& =- a^{(2)}_4 \left(a \cdot n \right)^2 \hat \upsilon^c - \left[s_4^{(2)} T + n^{(2)}_4 \left(\mu + \frac{1}{2} \bar u^2 \right) \right] \left(a \cdot n \right)^2 u^c  \\
&- 4 a^{(2)}_4 \left(a \cdot n \right)^2  u^c , \nonumber \\
    \left(\mathcal{T}_S \right)^{cd}_{S_{(4)}} & = a^{(2)}_4 \left(a \cdot n \right)^2 h^{cd} + n^{(2)}_4 \left(a \cdot n \right)^2 u^c u^d,  \\
    \mathcal{D}^{cd}_{S_{(4)}} & = 2 \left(a \cdot n \right) a^{(2)}_4 u^c u^d . 
\end{align}
%%%%%
The variation of $S_{(5)}=a^c a^d \bar h_{cd}$ is
%%%%
\begin{align}
    \delta S_{(5)} = 2 \delta a^c a^d \bar h_{cd} + a^c a^d \delta \bar h_{cd},
\end{align}
%%%%
contributing to the variation of the equilibrium partition function as
%%%%
\begin{align}
    \delta \mathcal{F}_{S_{(5)}} &= \frac{a^{(2)}_5}{2} \int d^{d-1} \sigma \hat e \left[ \frac{ \xi_5^{(2)cd} S_{(5)}}{a^{(2)}_5} + 2 a^c a^d + 4a_f h^{cf} h^{dg} D_d \mu \right. \\
    & \left. + 2  h^{fg}D_f \bar a_g u^c u^d \right]\delta \bar h_{cd} \nonumber \\ 
    & - \int d^{d-1} \sigma \hat e \left[ \left( \zeta_5^{(2)c} S_{(5)} + 4 a^{(2)}_5 \hat \upsilon^{(f}h^{g) c} \bar a_f D_g \mu \right. \right. \nonumber \\
    & \left. \left.+  2 a^{(2)}_5 u^c \left(\mu + \frac{1}{2} \bar u^2 \right) h^{fg} D_fD_g\mu  \right)\delta \tau_c \right]. \nonumber
\end{align}
%%%%
{The contribution of this variation to the currents is given by
%%%%
\begin{align}
    \left(\mathcal{T}_S\right)^c_{S_{(5)}}& =- a^{(2)}_5 \bar a^2 \hat \upsilon^c - \left[s_5^{(2)} T + n^{(2)}_5 \left(\mu + \frac{1}{2} \bar u^2 \right) \right] \bar a^2 u^c  \\
    \nonumber 
    &   - 4 a^{(2)}_5 \hat \upsilon^{(f}h^{g)c}a_f D_g \mu- 2 a^{(2)}_5 u^c \left( \mu + \frac{1}{2}\bar u^2 \right) h^{fg} D_f D_g \mu ,  \\
    \left(\mathcal{T}_S \right)^{cd}_{S_{(5)}}& = a^{(2)}_5 \bar a^2 h^{cd} + n^{(2)}_5 \bar a^2 u^c u^d + 2 a^{(2)}_5 a^c a^d + 4 a^{(2)}_5 \bar a_f h^{f(c} h^{d)g} D_g \mu \\
    &+ 2 a^{(2)}_5 h^{fg} \left(D_f \bar a_g\right) u^c u^d , \nonumber \\
    \mathcal{D}^{cd}_{S_{(5)}} &  = 0 .  
\end{align}
%%%%%
The variation of $S_{(6)}=K_{cd}u^c a^d$ is
%%%
\begin{align}
    \delta S_{(6)}=\left(a^c u^d \right) \delta K_{cd} - \left(K_{fg} a^a u^g u^c \right) \delta \tau_c + K_{cd} u^c \delta a^d,
\end{align}
contributing to the variation of the equilibrium partition function as
%%%%
\begin{align}
    \delta \mathcal{F}_{S_{(6)}} &=\int \frac{d^{d-1} \sigma \hat e}{2}    \left[ \xi^{(2)cd}_6 S_{(6)} + a^{(2)}_{6}\left( 2  K_{ab} u^a h^{bc} h^{fd} D_f \mu \right. \right. \\
    & \left. \left.+h^{bf} D_f\left( K_{ab}u^a \right) u^c u^d \right)  \right] \delta \bar h_{cd} \nonumber \\ 
    \nonumber &-  \int d^{d-1} \sigma \hat e \left(\zeta^{(2)c}_6 S_{(6)} + a^{(2)}_{6} K_{fg} a^f u^g u^c + 2a^{(2)}_{6} K_{ab} u^a \hat \upsilon^{(b}h^{f)c} D_f \mu \right)\delta \tau_c\\
    \nonumber &+\int d^{d-1} \sigma \hat e \left[ - h^{bf} D_f   \left( a^{(2)}_6 K_{ab}u^a \right) \left( \mu + \frac{1}{2} \bar u^2  \right)u^c \delta \tau_c \right. \nonumber \\
    & \left. +  \left(a^{(2)}_6 a^c u ^d \right)\delta K_{cd}  \right] . \nonumber
\end{align}
%%%%
%%%%
The contribution of this variation to the currents is given by
%%%%
\begin{align}
    \left(\mathcal{T}_S\right)^c_{S_{(6)}}& =- a^{(2)}_6 \left( \bar a \cdot \left(n \cdot \Omega\right) \right) \hat \upsilon^c - \left[s_6^{(2)} T + n^{(2)}_6 \left(\mu + \frac{1}{2} \bar u^2 \right) \right] \left( \bar a \cdot \left(n \cdot \Omega\right) \right) u^c 
    \nonumber \\
    &- a^{(2)}_6 K_{fg} a^f u^g u^c   - 2 a^{(2)}_6 K_{ab} u^a \hat \upsilon^{(b} h^{f)c} D_f \mu 
    \nonumber \\
    &- h^{bf}D_f \left(a^{(2)}_6 K_{ab}u^b \right) \left(\mu + \frac{1}{2} \bar u^2 \right) u^c ,  \\
    \left(\mathcal{T}_S \right)^{cd}_{S_{(6)}} & = a^{(2)}_6 \left( \bar a \cdot \left(n \cdot \Omega\right) \right) h^{cd} + n^{(2)}_6 \left( \bar a \cdot \left(n \cdot \Omega\right) \right) u^c u^d \nonumber \\
    &+  2 a^{(2)}_6 K_{ab} u^a h^{bc} h^{fd} D_f \mu + a^{(2)}_6 h^{bf} D_f\left(K_{ab} u^a \right)u^c u^d ,  \\
    \mathcal{D}^{cd}_{S_{(6)}} &  = a^{(2)}_6 a^{(c} u^{d)} .
\end{align}
%%%%
The variation of $S_{(7)}=\Omega^{ab}\Omega^{cd}\bar h_{ac} \bar h_{bd}$ is 
%%%%
\begin{align}
    \delta S_{(7)} &= 2 \bar h_{ab} \Omega^{ac} \Omega^{bd} \delta \bar h_{cd} + 2 \bar h_{ac} \bar h_{bd} \Omega^{ab} \delta\Omega^{cd} \\
    &= \left[\bar u_a \bar h_{b c} \Omega^{ab} -D_c \left( \mu +\frac{1}{2} \bar u^2 \right)   \right] \Omega^{cd} \delta \tau_d  -  \left( \Omega^{af} \Omega^{bg} \bar h_{ab} \bar h_{fg} \right) u^c \delta \tau_c  \nonumber \\ \nonumber & + \frac{1}{2} \delta \bar h_{cd} \Omega^{cf} \Omega^{dg} \bar h_{fg} -  \bar h_{bc} \Omega^{cf} u^b u^d D_f \delta \tau_d  +  \Omega^{cd} u^b D_c \delta \bar h_{bd} ,  
\end{align}
%%%
contributing to the variation of the equilibrium partition function as
%%%
\begin{align}
\delta \mathcal{F}_{S_{(7)}} &=   \int \frac{d^{d-1} \sigma \hat e}{2}    \left[ \xi^{(2)cd}_7 S_{(7)} + 2 a^{(2)}_7 \Omega^{cf} \Omega^{dg} \bar h_{fg} - 2 D_b \left(a^{(2)}_{(7)} \Omega^{bd} u^c  \right)  \right] \delta \bar h_{cd} \\ \nonumber &-  \int d^{d-1} \sigma \hat e \left[\zeta^{(2)d}_7 S_{(7)} - a^{(2)}_{7} \bar u_a \bar h_{bc} \Omega^{ab} \Omega^{cd} \right. \nonumber \\  
& \left. + a^{(2)}_7 D_c \left( \mu + \frac{1}{2}\bar u^2 \right) \Omega^{cd} \right]\delta \tau_d \nonumber \\
    \nonumber &+\int d^{d-1} \sigma \hat e \left[ D_f \left(a^{(2)}_{7} \bar h_{bd} \Omega^{cd} u^b  \right) - a^{(2)}_{7} \left( \Omega^{af} \Omega^{bg} \bar h_{ab} \bar h_{fg} \right)    \right] u^c \delta \tau_c .
\end{align}
%%%%
%%%%
The contribution of this variation to the currents is given by
%%%%
\begin{align}
    \left(\mathcal{T}_S\right)^c_{S_{(7)}}& =- a^{(2)}_7 \bar \Omega^2 \hat \upsilon^c - \left[s_6^{(7)} T + n^{(2)}_7 \left(\mu + \frac{1}{2} \bar u^2 \right) \right]\bar \Omega^2 u^c \\
    &+ a^{(2)}_7 \bar u_a \bar h_{bd} \Omega^{ab}\Omega^{dc}  - a^{(2)}_7 D_d\left( \mu + \frac{1}{2} \bar u^2 \right) \Omega^{dc} 
    \nonumber \\
    \nonumber &  + D_f\left(a^{(2)}_7 \bar h_{bd} \Omega^{df} u^b \right) u^c -a^{(2)}_7 \left( \Omega^{af}\Omega^{bg} \bar h_{ab} \bar h_{fg} \right) u^c,  \\
    \left(\mathcal{T}_S \right)^{cd}_{S_{(7)}} & = a^{(2)}_7 \bar \Omega^2 h^{cd} + n^{(2)}_7 \bar \Omega^2 u^c u^d - 2 a^{(2)}_7 \Omega^{f(c}\Omega^{d)g} \bar h_{fg}- 2 D_b \left(a^{(2)}_7 \Omega^{b(c} u^{d)} \right),  \\
    \mathcal{D}^{cd}_{S_{(7)}}  &  = 0. 
\end{align}

\section{Appendix}
\label{Appendix}

Figure \ref{fig:PlotBarKappa} shows the plot of the aspect ratio $\mathcal{A}$ of numerical solutions to (\ref{YoungLaplaceFunctional}), vs the dimensionless vorticity $\hat{\omega}_{\bar{\kappa}}^2$, which can take negative values.
Similar to figure \ref{fig:PlotKappa}, horizontal dashed lines determine intervals of $\mathcal{A}$ that correspond to different characteristic shapes, the same ones shown in figure \ref{fig:PlotKappa}.

\begin{center}
\begin{figure}[h!]
    \centering
    \includegraphics[scale=0.4]{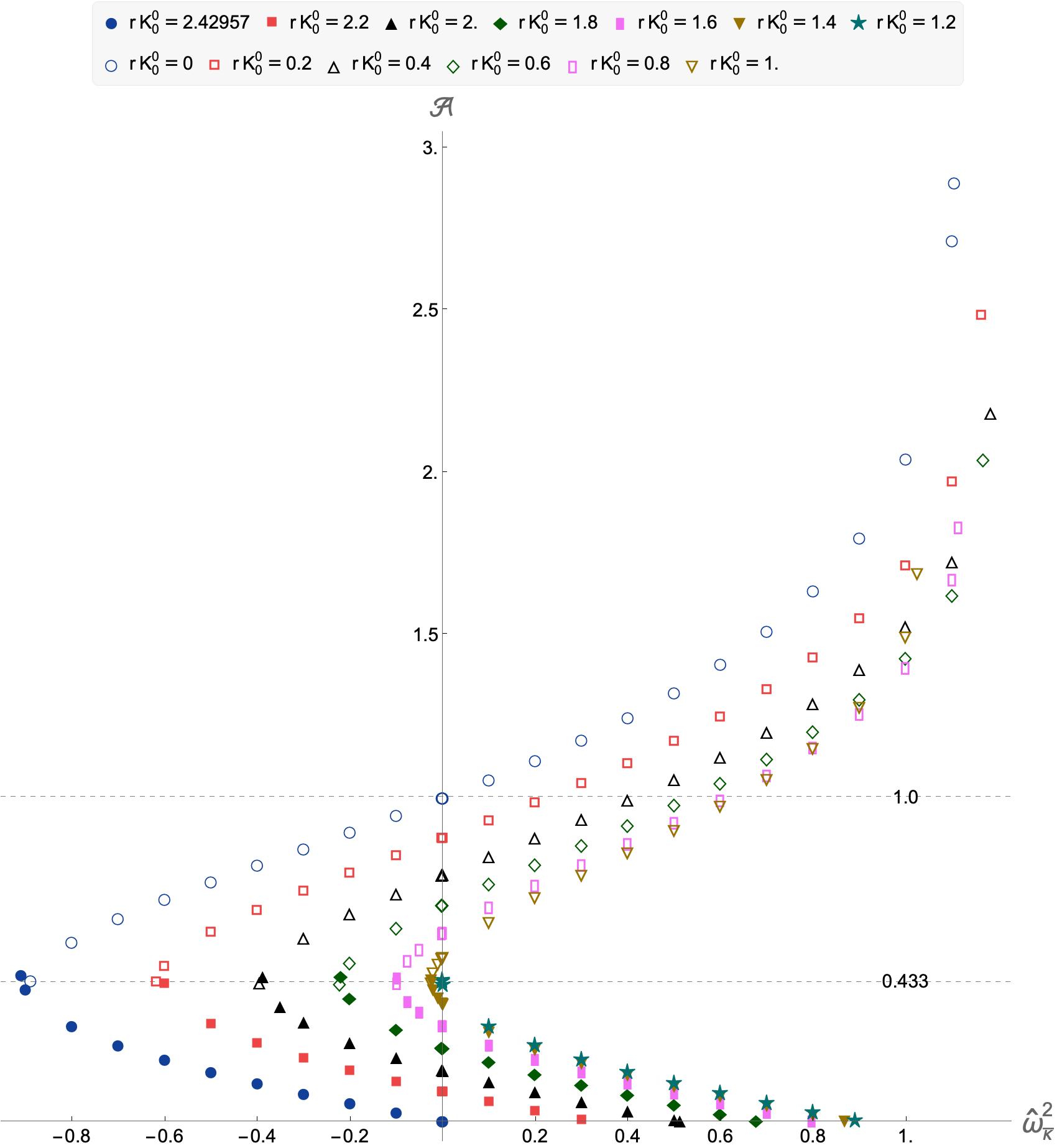}
    \caption{
    Plot of $\mathcal{A}$ vs $\hat{\omega}_{\bar{\kappa}}^2$ for a series of values of $rK_0^{(0)}$.
    }
    \label{fig:PlotBarKappa}
\end{figure}
\end{center}

%\nocite{*}
\bibliographystyle{plain}
\bibliography{apssamp}% Produces the bibliography via BibTeX.

\end{document}